\documentclass[11pt]{article}

\usepackage[final]{acl}

\usepackage{times}
\usepackage{latexsym}

\usepackage[T1]{fontenc}
\usepackage[utf8]{inputenc}

\usepackage{microtype}

\usepackage{inconsolata}

\usepackage{graphicx}

\usepackage{graphicx}
\usepackage{subfig}
\usepackage{booktabs} % 用于专业表格线
\usepackage{multirow} % 用于合并单元格
\usepackage{array}    % 用于列格式调整
\usepackage{makecell} % 用于单元格内换行
\usepackage{enumitem}
\usepackage{wrapfig}
\usepackage{amsmath}    % For \text command
\usepackage{microtype} % 改善字距
\usepackage{caption}  % 必须
\usepackage{subcaption}

\usepackage{times}
\usepackage{latexsym}
\usepackage{enumitem}
\usepackage{comment}

\usepackage{multirow}
\usepackage{pdflscape} % rotate page to landscape
\usepackage{adjustbox} % rotate area
\usepackage{longtable} % for the long table that is broken into several pages
\usepackage{hyperref} % for hyperlinks
\usepackage[normalem]{ulem}  % 
\useunder{\uline}{\ul}{}
\usepackage{array}
\newcolumntype{L}[1]{>{\raggedright\arraybackslash}p{#1}}

\usepackage{booktabs}
\usepackage{amsmath}
\usepackage{makecell} 

\makeatletter
\renewcommand{\subparagraph}{%
  \@startsection{subparagraph}{5}{\z@}{1ex \@plus 1ex \@minus .2ex}{-1em}{\normalfont\normalsize\itshape}}
\makeatother

\usepackage{amsthm} % For theorem-like environments
\usepackage{mdframed} % Include the mdframed package
\newtheoremstyle{italicstyle}
  {0em}   % Space above
  {0.5em}   % Space below
  {\itshape}  % Body font
  {}      % Indent amount (none)
  {\itshape} % Theorem head font
  {.}     % Punctuation after theorem head
  {\newline}  % Space after theorem head (space to insert after the head, before the text)
  {}  % Theorem head spec, use newline and set body indent to 2em

\theoremstyle{italicstyle}
\newtheorem{exampleinner}{Example}

\usepackage{setspace}  % For adjusting line spacing

\usepackage{hhline}

\usepackage[edges]{forest}
\usepackage{tikz}

\usepackage[ruled,vlined]{algorithm2e}

\usepackage{amsfonts}
\usepackage{amssymb}

\usepackage{pifont}% http://ctan.org/pkg/pifont
\newcommand{\cmark}{\ding{51}}%
\newcommand{\xmark}{\ding{55}}%

\begin{document}

% \title{From Raw Edits to Cognitive Intentions: A Comprehensive Survey of Edit Intention Data, Methods, and Models}
%\title{Making Revisions Understandable: A Comprehensive Survey of Edit Intention Data, Methods, and Models}
%\title{Making Revisions Understandable: A Survey of Edit Intention Data, Methods, and Models}
\title{Making Revisions Understandable: A Survey of Edit Intentions, Methods, and Applications}
%\title{Making Revisions Understandable: A Survey of Edit Intentions in Text Revision}
%\title{A Survey of Edit Intention Data, Methods, and Applications}

% %%
% %% The "author" command and its associated commands are used to define
% %% the authors and their affiliations.
% %% Of note is the shared affiliation of the first two authors, and the
% %% "authornote" and "authornotemark" commands
% %% used to denote shared contribution to the research.
% \author{Fangping Lan}
% \affiliation{%
%   \institution{Temple University}
%   \city{Philadelphia}
%   \state{Pennsylvania}
% \email{fangping.lan@temple.edu}

% \author{Qi Zhang}
% \affiliation{%
%   \institution{Temple University}
%   \city{Philadelphia}
%   \state{Pennsylvania}
% \email{qi.zhang@temple.edu}

% \author{Eduard Dragut}
% \affiliation{%
%   \institution{Temple University}
%   \city{Philadelphia}
%   \state{Pennsylvania}
\author{
 \textbf{Fangping Lan} \quad
 \textbf{Qi Zhang} \quad
 \textbf{Eduard C. Dragut}
\\
Temple University
\\
\texttt{\{fangping.lan, qi.zhang, edragut\}@temple.edu}
}

\maketitle

\begin{abstract}

% %%%%%%%%%%%%%%% Long paper version begin %%%%%%%%%%%%%%%
% \input{sections/abstract}
% %%%%%%%%%%%%%%%% Long paper version end %%%%%%%%%%%%%%%

%%%%%%%%%%%%%%% ACL2026 version begin %%%%%%%%%%%%%%%
Text revision is a core process in document creation, capturing how authors iteratively refine, reorganize, and improve written content. With the increasing availability of large-scale revision histories from platforms such as Wikipedia and arXiv, NLP research has begun to move beyond modeling what changes are made to understanding why they are made, i.e., the underlying edit intentions. 
To our knowledge, this is the \emph{first} survey\footnote{The GitHub repository is available at \url{https://github.com/TUDMLab/MakeRevisionsUnderstandable}} that synthesizes text revision research through the lens of edit intentions, providing a unified view of datasets, taxonomies, identification methods, and applications. 
We review prior work across the full revision workflow, including revision corpus construction, edit intention taxonomy design, and edit intention identification. We further categorize representative datasets and methods, summarize downstream applications such as writing assistance and document edit summarization, and highlight key open research directions.
%%%%%%%%%%%%%%%% ACL2026 version end %%%%%%%%%%%%%%%

\end{abstract}

% %%%%%%%%%%%%%%% Long paper version begin %%%%%%%%%%%%%%%
% \input{sections/introduction}
% \input{sections/definitions}
% \input{sections/datasets-ed}
% \input{sections/revision_extraction-ed}
% \input{sections/edit_intention_categorization}
% \input{sections/methods}
% \input{sections/application-ed}
% \input{sections/discussion-ed}
% \input{sections/limitations-ed}
% %%%%%%%%%%%%%%%% Long paper version end %%%%%%%%%%%%%%%

%%%%%%%%%%%%%%% ACL2026 version begin %%%%%%%%%%%%%%%

\section{Introduction}

\tikzstyle{my-box}=[
    rectangle,
    draw=gray,
    rounded corners,
    text opacity=1,
    minimum height=1.5em,
    minimum width=5em,
    inner sep=2pt,
    align=center,
    fill opacity=.5,
    line width=0.8pt,
]
\tikzstyle{leaf}=[my-box, minimum height=1.5em,
    fill=pink!10, text=black, align=left,font=\normalsize,
    inner xsep=2pt,
    inner ysep=4pt,
    line width=0.8pt,
]

\definecolor{c1}{RGB}{93,191,237} % blue
\definecolor{c2}{RGB}{237,110,106} % red
\definecolor{c3}{RGB}{240,154,69} % yellow
\definecolor{c4}{RGB}{8,153,68} % green
\definecolor{c5}{RGB}{205,180,243} % purple
\definecolor{c6}{RGB}{97,218,184} % cyan
\definecolor{c7}{RGB}{0,128,128} % teal

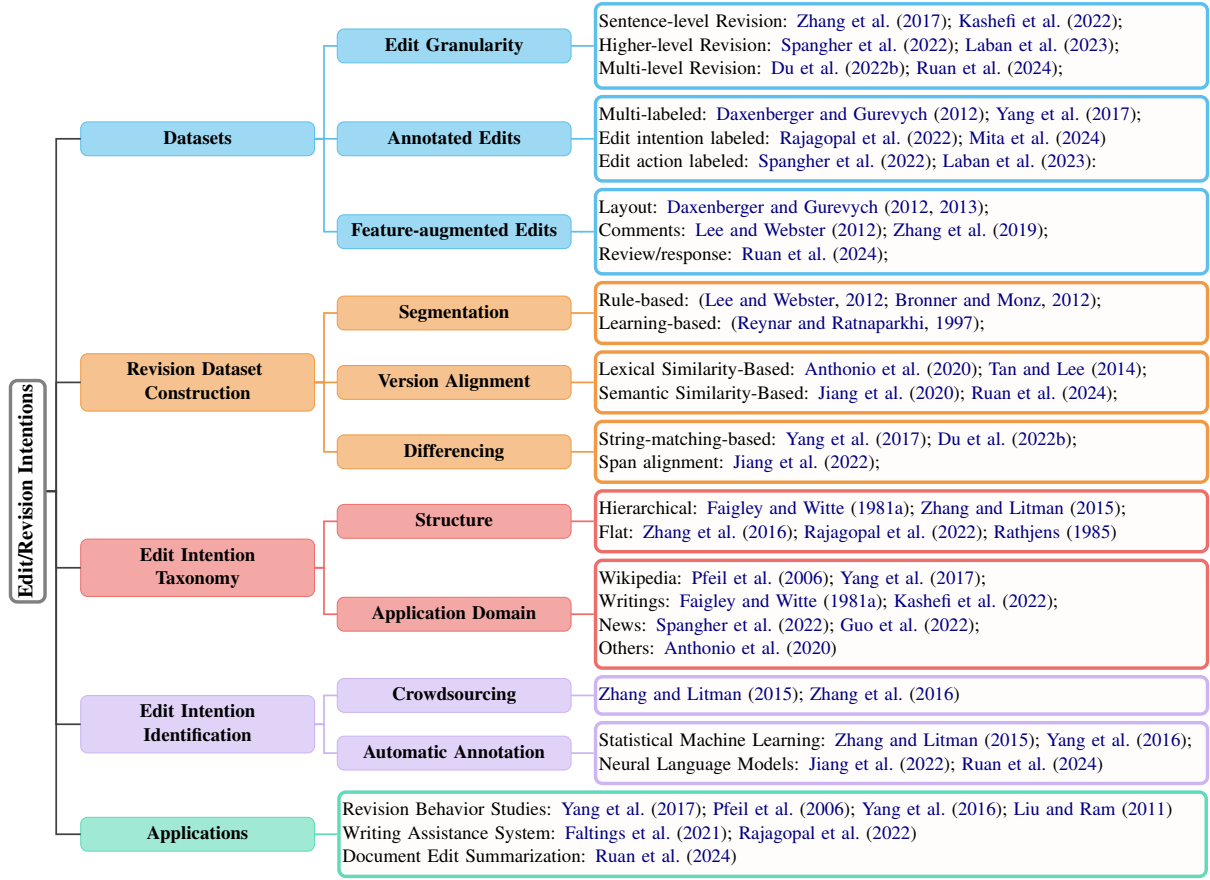
\begin{figure*}[!t]
    \centering
    \resizebox{\textwidth}{!}{
        \begin{forest}
            forked edges,
            for tree={
                grow=east,
                reversed=true,
                anchor=base west,
                parent anchor=east,
                child anchor=west,
                base=center,
                font=\large,
                rectangle,
                % draw=hidden-draw,
                draw=gray,
                rounded corners,
                align=left,
                text centered,
                minimum width=4em,
                edge+={darkgray, line width=1pt},
                s sep=3pt,
                inner xsep=2pt,
                inner ysep=3pt,
                line width=0.8pt,
                ver/.style={rotate=90, child anchor=north, parent anchor=south, anchor=center},
            },
            where level=1{text width=12em,font=\normalsize,}{},
            where level=2{text width=12em,font=\normalsize,}{},
            where level=3{text width=10em,font=\normalsize,}{},
            where level=4{text width=35em,font=\normalsize,}{},
            where level=5{text width=18em,font=\normalsize,}{},
            [
                \textbf{Edit/Revision Intentions}, ver, line width=0.7mm
                [
                    \textbf{Datasets}, fill=c1!60, draw=c1, line width=0mm
                    [   
                        \textbf{Edit Granularity}, fill=c1!60, draw=c1, line width=0mm, edge={c1}
                        [
                            Sentence-level Revision: \citet{zhang-etal-2017-corpus, Kashefi2022-xs}; \\
                            Higher-level Revision: \citet{spangher-etal-2022-newsedits, laban-etal-2023-swipe}; \\
                            Multi-level Revision: \citet{du-etal-2022-understanding-iterative, ruan-etal-2024-re3};
                            , leaf, text width=32em, draw=c1, line width=0.7mm, edge={c1}
                        ]
                    ]
                    [
                        \textbf{Annotated Edits}, fill=c1!60, draw=c1, line width=0mm, edge={c1}
                        [
                              Multi-labeled: \citet{daxenberger-gurevych-2012-corpus, yang-etal-2017-identifying-semantic};\\
                              Edit intention labeled: \citet{rajagopal-etal-2022-one, mita-etal-2024-towards} \\
                              Edit action labeled: \citet{spangher-etal-2022-newsedits, laban-etal-2023-swipe}: \\
                              , leaf, text width=32em, draw=c1, line width=0.7mm, edge={c1}
                        ]
                    ]
                    [
                        \textbf{Feature-augmented Edits}, fill=c1!60, draw=c1, line width=0mm, edge={c1}
                        [
                              Layout: \citet{daxenberger-gurevych-2012-corpus, daxenberger-gurevych-2013-automatically}; \\ 
                              Comments: \citet{lee-webster-2012-corpus, zhang-etal-2019-modeling};  \\
                              Review/response: \citet{ruan-etal-2024-re3}; \\
                              % Multilingual: \citet{faruqui-etal-2018-wikiatomicedits, spangher-etal-2022-newsedits}; 
                              ,leaf, text width=32em, draw=c1, line width=0.7mm, edge={c1}
                        ]
                    ]
                ]% level1
                % [
                %     \textbf{Segmentation}, align=center,
                %     fill=c2!60, draw=c2, line width=0mm
                %     [
                %         Dynamic Programming; Regex Rules; CRF model; Tokenizer ,leaf, text width=32em, draw=c2, line width=0.7mm, edge={c2}
                %     ]
                % ]% level1
                [
                    \textbf{Revision Dataset } \\\textbf{Construction}, align=center, fill=c3!60, draw=c3, line width=0mm
                    [
                        \textbf{Segmentation}, fill=c3!60, draw=c3, line width=0mm, edge={c3}
                        [
                         Rule-based: \cite{lee-webster-2012-corpus, bronner-monz-2012-user}; \\
                         Learning-based: \cite{reynar-ratnaparkhi-1997-maximum}; \\
                        , leaf, text width=32em, draw=c3, line width=0.7mm, edge={c3}
                        ]
                    ]
                    [
                        \textbf{Version Alignment}, fill=c3!60, draw=c3, line width=0mm, edge={c3}
                        [
                         Lexical Similarity-Based: \citet{ anthonio-etal-2020-wikihowtoimprove, tan-lee-2014-corpus}; \\ 
                         Semantic Similarity-Based: \citet{jiang-etal-2020-neural, ruan-etal-2024-re3};\\
                        , leaf, text width=32em, draw=c3, line width=0.7mm, edge={c3}
                        ]
                    ]
                    [
                        \textbf{Differencing}, fill=c3!60, draw=c3, line width=0mm, edge={c3}
                        [
                        String-matching-based: \citet{yang-etal-2017-identifying-semantic, du-etal-2022-understanding-iterative};\\
                        Span alignment: \citet{jiang-etal-2022-arxivedits}; \\
                        % Section-title based: \citet{tan-lee-2014-corpus}; \\
                        , leaf, text width=32em, draw=c3, line width=0.7mm, edge={c3}
                        ]
                    ]
                ]% level1
                [
                    \textbf{Edit Intention} \\\textbf{Taxonomy}, align=center, fill=c2!60, draw=c2, line width=0mm
                    [
                        \textbf{Structure},  fill=c2!60, draw=c2, line width=0mm, edge={c2}
                        [
                        Hierarchical: \citet{Faigley1981, zhang-litman-2015-annotation}; \\
                        Flat: \citet{zhang-etal-2016-argrewrite, rajagopal-etal-2022-one, Rathjens1985}
                        , leaf, text width=32em, draw=c2, line width=0.7mm, edge={c2}
                        ]
                    ]
                    [
                        \textbf{Application Domain},  fill=c2!60, draw=c2, line width=0mm, edge={c2}
                        [
                         Wikipedia: \citet{Pfeil2006, yang-etal-2017-identifying-semantic};\\
                         Writings: \citet{Faigley1981, Kashefi2022-xs}; \\
                         News: \citet{spangher-etal-2022-newsedits, Guo2022}; \\
                         Others: \citet{anthonio-etal-2020-wikihowtoimprove}
                        , leaf, text width=32em, draw=c2, line width=0.7mm, edge={c2}
                        ]
                    ]
                    % [
                    %     \textbf{Lineage-aware},  fill=c2!60, draw=c2, line width=0mm, edge={c2}
                    %     [
                    %      \citet{lan-etal-2025-unit, zhang-etal-2016-argrewrite, jiang-etal-2022-arxivedits}\\ 
                    %     , leaf, text width=32em, draw=c2, line width=0.7mm, edge={c2}
                    %     ]
                    % ]
                ]% level1
                [
                    \textbf{Edit Intention}\\\textbf{Identification}, align=center, fill=c5!60, draw=c5, line width=0mm
                    [ 
                        \textbf{Crowdsourcing}, fill=c5!60, draw=c5, line width=0mm, edge={c5}
                        [
                        \citet{zhang-litman-2015-annotation, zhang-etal-2016-argrewrite} 
                        , leaf, text width=32em, draw=c5, line width=0.7mm, edge={c5}
                        ]
                    ]
                    [ 
                        \textbf{Automatic Annotation}, align=center, fill=c5!60, draw=c5, line width=0mm, edge={c5}
                        [
                        Statistical Machine Learning: \citet{zhang-litman-2015-annotation, Yang_Halfaker_Kraut_Hovy_2016}; \\
                        Neural Language Models: \citet{jiang-etal-2022-arxivedits, ruan-etal-2024-re3}
                        , leaf, text width=32em, draw=c5, line width=0.7mm, edge={c5}
                        ]
                    ]
                ]% level1
                [
                    \textbf{Applications}, fill=c6!60, draw=c6, line width=0mm
                    [   
                        Revision Behavior Studies: \citet{yang-etal-2017-identifying-semantic, Pfeil2006, Yang_Halfaker_Kraut_Hovy_2016, Liu2011} \\
                        Writing Assistance System: \citet{faltings-etal-2021-text, rajagopal-etal-2022-one} \\
                        Document Edit Summarization: \citet{ruan-etal-2024-re3}
                        % \textbf{Revision Behavior Studies},  fill=c6!60, draw=c6, line width=0mm, edge={c6}
                        % [
                        % Interactive Text Editing \cite{faltings-etal-2021-text}; \\
                        % Iterative Text Editing \cite{du-etal-2022-understanding-iterative}; 
                        , leaf, text width=45.5em, draw=c6, line width=0.7mm, edge={c6}
                        ]
                    ]
                ]
            ]
        \end{forest}
    }
    % \vspace{-20pt}
    \caption{Taxonomy of edit intention-related research. We only list representative works for each kind of task, and for a more complete version, please refer to Appendix \ref{sec:complete_research_taxonomy}).}
    \label{fig:taxonomy}
    % \vspace{-15pt}
\end{figure*}

Text revision is a fundamental process in document creation, reflecting how authors iteratively refine content, correct errors, and reshape meaning over time. Unlike final drafts, revisions preserve explicit traces of authors’ decision-making, making them a valuable source for studying writing behavior and textual evolution \citep{Faigley1981}. The widespread availability of large-scale revision histories from platforms such as Wikipedia \cite{daxenberger-gurevych-2012-corpus} and arXiv \cite{du-etal-2022-read, jiang-etal-2022-arxivedits} has further enabled empirical study of revision processes in naturalistic settings \cite{zhang-etal-2017-corpus, Kashefi2022-xs}.

% Beyond identifying what textual changes occur, understanding \emph{why} authors make these changes—referred to as \emph{edit intentions}—is crucial for interpreting revisions at a semantic and pragmatic level\cite{yang-etal-2017-identifying-semantic}. Edit intentions provide an abstraction over surface-level diffs \cite{zhang-etal-2016-argrewrite}, capturing the underlying goals of revisions such as content elaboration, factual correction, stylistic refinement, or structural reorganization \cite{spangher-etal-2022-newsedits}. This perspective has supported a range of NLP applications, including revision behavior analysis \cite{Pfeil2006}, writing assistance \cite{zhang-litman-2015-annotation}, text simplification \cite{laban-etal-2023-swipe}, and document edit summarization \cite{ruan-etal-2024-re3}.

Beyond identifying \emph{what} textual changes occur, understanding \emph{why} authors make these changes, referred to as \emph{edit intentions}, is crucial for interpreting revisions at a semantic and pragmatic level \cite{yang-etal-2017-identifying-semantic}. Edit intentions provide a high-level abstraction over surface-level differences \cite{zhang-etal-2016-argrewrite}, capturing the underlying goals of revisions such as content elaboration, factual correction, stylistic refinement, or structural reorganization \cite{spangher-etal-2022-newsedits}. This perspective has catalyzed a range of NLP applications, including writing assistance \cite{zhang-litman-2015-annotation}, text simplification \cite{laban-etal-2023-swipe}, and document edit summarization \cite{ruan-etal-2024-re3}. Beyond revision settings, recent work shows that intent taxonomies can generalize to other forms of content interaction, such as modeling the communicative role of hyperlinks in social media posts \cite{lan2026linkintenttaxonomyincluding}.

% Research on edit intention spans multiple interconnected components, including revision processing (e.g., segmentation, version alignment, and differencing), edit intention taxonomy design, and edit intention identification. However, existing work remains fragmented across datasets, domains, taxonomies, and modeling assumptions. Edit intention taxonomies vary widely in structure and scope, and identification methods range from manual annotation to feature-based models and large language models (LLMs), often evaluated under incompatible settings.

% Despite steady progress, to our knowledge, there is currently no comprehensive survey that synthesizes text revision research through the unified lens of edit intentions. This survey fills that gap by systematically reviewing datasets, methods, taxonomies, and applications related to edit intention modeling, while highlighting shared challenges and open research directions.

% Research on edit intentions spans multiple components of the revision pipeline—revision dataset construction \cite{zhang-litman-2014-sentence}, edit intention taxonomy (EIT) design \cite{lan-etal-2025-unit}, and edit intention identification \cite{daxenberger-gurevych-2013-automatically}—yet the literature remains fragmented across domains, taxonomies, and evaluation settings. 

Prior works cover writing processes, collaborative editing, and text generation, but none treat \emph{edit intentions} as a first-class object across the revision workflow.
Early composition studies distinguish surface-level versus meaning-changing edits yet predate computational models, large-scale corpora, and standardized evaluation protocols \citep{Sommers1980, Faigley1981}. 
Surveys of collaborative writing (e.g., Wikipedia) primarily examine editor behavior and quality control, using edit types mainly as auxiliary signals rather than modeling intentions \citep{Pfeil2006, Halfaker2013}. More recent NLP surveys emphasize \emph{single-shot} transformations such as style transfer or simplification, without considering revision histories, alignment and differencing, or taxonomy design \citep{jin-etal-2022-deep}. Meanwhile, dataset- and method-specific studies introduce corpora or models for edit intention identification or revision summarization, but do not synthesize shared design choices, annotation practices, or methodological trade-offs across domains and settings \citep{zhang-litman-2015-annotation, spangher-etal-2022-newsedits, ruan-etal-2024-re3}.

Thus, the literature remains fragmented along several recurring axes. 
First, choices in revision workflows (e.g., segmentation granularity, version alignment, and differencing) directly determine what can be reliably labeled and learned downstream.
Second, edit intention taxonomies (EITs) are often domain-specific and defined with varying levels of granularity and scope, hindering cross-corpus comparability.
Third, identification methods and evaluations are frequently tied to particular label sets and protocols, limiting reproducibility and cross-domain generalization.
To address these challenges, we adopt an \emph{edit intention--centric} perspective that integrates revision corpus construction, EIT design, edit intention identification, and downstream applications under a unified view. To our knowledge, this is the \emph{first}  survey that synthesizes text revision research explicitly through the lens of edit intentions. We consolidate terminology and systematically organize prior work across datasets, corpus construction workflows, taxonomy design, identification methods, applications, and shared open challenges. We first identified seed papers using keyword-based searches (e.g., ``edit intention'', ``revision taxonomy'', ``writing revision'') across major digital libraries. We then performed backward and forward snowballing by examining references and citing papers to expand coverage. We included papers that either (1) propose, refine, or analyze an EIT, or (2) develop downstream tasks that explicitly leverage edit intentions. We iterated this process until no additional eligible works were identified.

% \noindent\textbf{Survey Scope.} 
This survey aims to provide a unified and structured view of text revision research through the lens of edit intentions. Specifically, we:
(i) establish consistent terminology for revision-related concepts;
(ii) systematically categorize datasets and methods across the revision pipeline;
(iii) analyze approaches to edit intention taxonomy construction and identification;
(iv) summarize empirical findings across key application domains; and
(v) identify promising open directions for future research.
% \noindent\textbf{Survey Contributions.}
In summary, our contributions are:
\begin{itemize}[
  leftmargin=*,        % align list with surrounding text
  labelsep=0.35em,     % space between bullet and text (shrink this)
  itemsep=0pt,         % space between items
  parsep=0pt,
  topsep=2pt,
  partopsep=0pt
]
\item \textbf{Unified framework.} We consolidate terminology and organize prior work across  (1) the full revision workflow, (2) datasets and corpus construction, (3) EITs, (4) identification methods, (5) applications, and (6) open challenges, into a unified \emph{six-dimensional} view. Figure~\ref{fig:taxonomy} presents the first 5 views.
% \vspace{-3pt}
\item \textbf{Lineage-aware taxonomy analysis.} We introduce a \emph{lineage-aware} perspective to characterize how EITs evolve across domains and granularities, and summarize common operations such as merging, splitting, and refinement.
% \vspace{-3pt}
\item \textbf{Methods-to-applications map.} We review edit intention identification approaches spanning manual annotation, crowdsourcing, neural models, and LLM-based methods, and connect them to downstream uses including writing assistance, revision behavior analysis, and document edit summarization, highlighting methodological trade-offs and evaluation pitfalls.
\end{itemize}

\section{Formulation}
We introduce core definitions and notation to clarify key concepts in revision analysis and edit intention modeling and to support consistent comparison across datasets, methods, and applications.

%\subsection{Revision and Edits}
%We defined an edit as a single coherent change in a document. It formulates to the insertion, deletion, or substitution of a sub-expression $p$ such that both the original expression $s$ and the resulting expression $e(s)$ are well-formed semantic constituents \cite{MacCartney_2009, jiang-etal-2022-arxivedits}. E.g. $s$ =``She died from an illness'', $p$= ``in 1949'', and $e(s)$ = ``She died in 1949 from an illness''. This formulation is desirable because it exposes a relationship between the surface form and the semantics of natural language while remaining agnostic about the underlying semantic representation. That is, the difference in ``meaning'' between s and $e(s)$ is exactly the ``meaning'' of $p$ (in context), regardless of how that meaning is represented.

An \emph{edit} is a single, coherent change to a document, formalized as the insertion, deletion, or substitution of a sub-expression $p$ such that both the original text $s^{t-1}$ and the revised text $s^t = e(s^{t-1})$ are well-formed semantic constituents \cite{MacCartney_2009, jiang-etal-2022-arxivedits}. For example, inserting $p$ = “in 1949” into $s^{t-1}$ = “She died from an illness” yields $s^t$ = “She died in 1949 from an illness.” This formulation captures the semantic contribution of $p$ independent of the underlying representation.

% An \emph{edit} is a single, coherent change to a document, formalized as the insertion, deletion, or substitution of a sub-expression $p$ such that both the original text $s^{t-1}$ and the resulting text $s^t = e(s^{t-1})$ are well-formed semantic constituents \cite{MacCartney_2009, jiang-etal-2022-arxivedits}. 
% For example, let $s^{t-1}$ = ``She died from an illness,'' $p$ = ``in 1949,'' and $e(s^{t-1})$ = ``She died in 1949 from an illness.''
% This formulation explicitly captures the relationship between surface form and meaning while remaining agnostic to the choice of underlying semantic representation. In particular, the semantic difference between $s^{t-1}$ and $s^t$ is precisely the meaning contributed by $p$ in context, independent of how that meaning is formally represented.

%\paragraph{Revision} A revision is created when the editor saves the changes to the document \cite{Yang_Halfaker_Kraut_Hovy_2016, yang-etal-2017-identifying-semantic, du-etal-2022-understanding-iterative}. The granularity of revision can vary from sentence, paragraph, to document; thus, one revision may contain one or more edits. A revision at granularity $g$ of the original expression $g^{t-1}$ and the revised expression $g^{t}$ is denoted as $R^{t, g}$, where $t$ indicates the version of the expression and $g\in\{D, P, S\}$ denoting as document, paragraph, and sentence, respectively. 

A \emph{revision} occurs when an editor saves changes to a document \cite{Yang_Halfaker_Kraut_Hovy_2016, yang-etal-2017-identifying-semantic, du-etal-2022-understanding-iterative}. Revisions may occur at the sentence ($S$), paragraph ($P$), or document ($D$) level, and a single revision can contain multiple edits. We denote a revision at granularity $g \in \{D,P,S\}$ between versions $t-1$ and $t$ as $R^{t,g}$, relating the original expression $g^{t-1}$ to its revised form $g^{t}$.

% A \emph{revision} is created when an editor saves changes to a document \cite{Yang_Halfaker_Kraut_Hovy_2016, yang-etal-2017-identifying-semantic, du-etal-2022-understanding-iterative}. 
% The granularity of a revision vary from sentence to paragraph to document; consequently, a single revision may contain one or more edits. 
% $R^{t,g}$ denotes a revision at granularity $g$ between the original expression $g^{t-1}$ and the revised expression $g^{t}$, where $t$ indexes the revision version and $g \in \{D, P, S\}$ denotes document-, paragraph-, and sentence-level revisions, respectively.

A \emph{document-level revision} $R^{t,D}$ corresponds to a pair of document versions $(D^{t-1}, D^{t})$. 
It comprises $|R^{t,D}|$ paragraph-level revisions, denoted $\{R^{t,P}_{i}\}$, where $i$ indexes paragraph-level revisions. 
Each paragraph-level revision $R^{t,P}_{i}$ may in turn contain $|R^{t,P}_{i}|$ sentence-level revisions, denoted $\{R^{t,S}_{ij}\}$. 
Finally, each sentence-level revision $R^{t,S}_{ij}$ consists of one or more edits $\{e_{k}\}$.

%One \textit{document-level revision} $R^{t, D}$ is aligned with a pair of documents $(D^{t-1}, D^{t})$. It contains $|R^{t, D}|$ paragraph-level revisions, $R^{t, P}_{i}$, where $i$ is the index of paragraph-level revisions accordingly. Each paragraph-level revision may include $|R^{t, P}_{i}|$ sentence-level revisions, and each sentence-level revision, $R^{t, S}_{i}$, contains one or more edits, $e_{i}$. 

%\subsection{Edit Action and Intention}

%\paragraph{Edit Action} 
%An edit action \textbf{$a_{k}$} specifies how edits are made to certain text objects \cite{du-etal-2022-understanding-iterative}, e.g., insert, delete, merge, split, where $k$ is the index of a list of edit actions. 

An \emph{edit action} $a_k$ specifies the operation applied to a text object during an edit \cite{du-etal-2022-understanding-iterative}, such as \textit{insert}, \textit{delete}, \textit{merge}, or \textit{split}, where $k$ indexes the set of possible edit actions.

An \emph{edit intention} $I_k$ represents the editor’s underlying goal when performing a specific edit. We assume that each edit action $a_k$ is associated with a single edit intention $I_k$. Section~\ref{sec:eit_intention_categorization} discusses the categorization of edit intentions.

%\paragraph{Edit Intention} An edit intention \textbf{$I_{k}$} reflects the revising goal of the editor when making a certain edit. We assume each edit action $a_{k}$ will only be labeled with one edit intention $I_{k}$. We will further discuss the edit intention taxonomy in Section \ref{sec:eit_categorization}.

%\subsection{Version Alignment}

% Identifying revision pairs from two lengthy documents is critical and challenging, especially complicated by the expansive scope for comparison and the presence of recurring content\cite{jiang-etal-2020-neural}. 
%Version alignment aims to align two versions of the same text. Given an original document $D^{t-1}$ (paragraph $P^{t-1}$) of $m$ paragraphs (sentences) and a revised document $D^{t}$ (paragraph $P^{t}$) of $n$ paragraphs (sentences), for each paragraph $p_{i}^{t-1}$ (sentence $s_{i}^{t-1}$) in the original document (paragraph), we aim to find its corresponding paragraph $p_{j}^{t} = e(p_{i}^{t-1})$ (sentence $s_{j}^{t} = e(s_{i}^{t-1})$) in revised document (paragraph). 

\emph{Version alignment} establishes correspondences between two versions of the same text. Given an original document or paragraph at version $t-1$ and its revised version at version $t$, the goal is to map each paragraph or sentence in the original text to its corresponding unit in the revised text, if such a correspondence exists.

% \emph{Version alignment} aims to establish correspondences between two versions of the same text. 
% Given an original document $D^{t-1}$ (or paragraph $P^{t-1}$) consisting of $m$ paragraphs (or sentences) and a revised document $D^{t}$ (or paragraph $P^{t}$) consisting of $n$ paragraphs (or sentences), the goal is to identify, for each paragraph $p^{t-1}_i$ (or sentence $s^{t-1}_i$) in the original text, its corresponding paragraph $p^{t}_j = e(p^{t-1}_i)$ (or sentence $s^{t}_j = e(s^{t-1}_i)$) in the revised text.

Given an edit intention taxonomy $EIT = \{I_1, \ldots, I_k\}$ and an original-revised text pair $(g^{t-1}, g^{t})$ with $m$ edits, \emph{edit intention identification} aims to assign the most likely intention $I_i \in EIT$ to each edit $e_i$. At the sentence level, this is typically framed as a multi-class classification task, whereas at the paragraph and document levels it is modeled as multi-class multi-label classification, allowing multiple intentions per revision.

%\subsection{Edit Intention Identification}
% Finally, given an edit intention taxonomy $EIT = \{I_1, \ldots, I_k\}$ and a pair of original and revised texts $(g^{t-1}, g^{t})$ containing $m$ edits or revisions, the goal of \emph{edit intention identification} is to determine the most likely edit intention $I_i \in EIT$ for each edit $e_i$.
% At the sentence level, this task is typically formulated as a multi-class classification problem, assigning each edit to one of the $k$ edit intention categories. In contrast, at the paragraph and document levels, edit intention identification is naturally modeled as a multi-class multi-label classification problem, where multiple edit intention labels may be assigned to a single revision.

%Finally, given an edit intention taxonomy including a list of edit intentions $EIT=\{I_{1}, ..., I_{k}\}$, a pair of the original text $g^{t-1}$ and the revised text $g^{t}$ containing $m$ edits/revisions. For each edit $e_{i}^{k}$, determine a most-likely edit intention $I_{i}$ to the pair. For sentence-level revision, it is a multi-class classification that classifies the edit intention into one of $k$ edit intentions, while it is basically a multi-class multi-label classification that assigns multiple labels for each paragraph/document-level revision. 

% \paragraph{Manual Annotation}

% \paragraph{Multi-label Classification}
% Transforming to a single or multiple binary classifications.

% \paragraph{Multi-class Classification}

% \paragraph{Generative LM Labeling}

\section{Datasets}
\label{sec:datasets}

We review publicly released datasets for text revision and edit intention research, excluding datasets if they do not contain annotated edit actions or edit intentions. 
The remaining corpora are primarily derived from Wikipedia, academic writing, and student essays. 
In this section, we organize existing datasets along three dimensions: 
\textit{edit granularity}, which determines the level at which revisions are analyzed;
% and enables different types of behavioral insights; 
\textit{ground-truth labels}, which are essential for modeling edit actions and intentions; 
% and for supporting downstream applications; 
and 
\textit{feature augmentation}, which provides additional contextual signals useful for edit intention identification and revision modeling. 
Table~\ref{tab:datasets} summarizes the statistics of the collected datasets\footnote{While we aim to cover all publicly available datasets, we acknowledge the possibility of missing some relevant work.}.

\begin{table*}[t]
\centering
%\small
\footnotesize
\begin{tabular}{p{0.18\linewidth}|l|l|l|l|l|l|l}
% \hline
\Xhline{1pt}
\textbf{Paper} &
  \textbf{Gran.} &
  \textbf{\# Pairs} &
  \textbf{H.} &
  \textbf{I.} &
  \textbf{A.} &
  \textbf{Source} &
  % \textbf{Multi labels} &
  \textbf{Features} \\ \hline
  % \citet{daxenberger-gurevych-2012-corpus}
% % A Corpus-Based Study of Edit Categories in   Featured and Non-Featured Wikipedia Articles 
% &
%   S &
%   2K &
%   \cmark &
%   \cmark & 
%   \cmark &
%   Wikipedia &
%   % \cmark &
%   Including   layout \\ \hline
  
% Automatically Classifying Edit Categories in   Wikipedia Revisions &
\citet{daxenberger-gurevych-2013-automatically}$^{\blacklozenge}$ &
  S &
  2K &
  \cmark &
  \cmark &
  \cmark &
  Wikipedia &
  % \cmark &
  Including   layout \\ \hline
  % Identifying Semantic Edit Intentions from   Revisions in Wikipedia &
\citet{yang-etal-2017-identifying-semantic} &
  S &
  5.7K &
  \cmark &
  \cmark &
  \xmark &
  Wikipedia &
  % \cmark &
  Multiple   labels \\ \hline
  
% WikiAtomicEdits: A Multilingual Corpus of   Wikipedia Edits for Modeling Language and Discourse &
\citet{faruqui-etal-2018-wikiatomicedits} &
  S &
  43M &
  \cmark &
  \cmark &
  \xmark &
  Wikipedia &
  % \xmark &
  Multilingual \\ \hline

% % A   Corpus of Annotated Revisions for Studying Argumentative Writing &
% \citet{zhang-etal-2017-corpus} &
%   S &
%   180 &
%   \cmark &
%   Student   Essays &
%   \cmark &
%   \cmark &
%   % \xmark  &
%   Revision   history \\ % \hline
  % ArgRewrite V.2: an Annotated Argumentative   Revisions Corpus &
\citet{Kashefi2022-xs}$^{\blacklozenge}$ &
  S &
  3K &
  \cmark &
  \cmark &
  \cmark &
  Student   Essays &
  % \xmark &
  \makecell[l]{Revision   history, Multiple labels} \\ \hline

% % wikiHowToImprove: A Resource and Analyses on   Edits in Instructional Texts &
% \citet{anthonio-etal-2020-wikihowtoimprove} &
%   S &
%   2.7M &
%   \xmark &
%   WikiHow &
%   \xmark &
%   \xmark &
%   % \xmark &
%   / \\ %\hline
% % Towards Modeling Revision Requirements in   wikiHow Instructions &
% \citet{bhat-etal-2020-towards} &
%   S &
%   6.95M &
%   \xmark &
%   WikiHow &
%   \xmark &
%   \xmark &
%   % \xmark &
%   \makecell[l]{Unrevised sentences \\ included} \\ \hline
% % Understanding Iterative Revision from   Human-Written Text &
% \citet{du-etal-2022-understanding-iterative} &
%   S, P &
%   31K* &
%   \cmark &
%   \begin{tabular}[c]{@{}l@{}}ArXiv\\      Wikipedia\\      WikiNews\end{tabular} &
%   \cmark &
%   \cmark &
%   % \xmark &
%   \makecell[l]{Across domain; \\ Revision history} \\ % \hline
% % Read, Revise, Repeat: A System Demonstration for   Human-in-the-loop Iterative Text Revision &
% \citet{du-etal-2022-read} &
%   S, P &
%   367K  &
%   \cmark &
%   \begin{tabular}[c]{@{}l@{}}ArXiv\\      Wikipedia\\      WikiNews\end{tabular} &
%   \cmark &
%   \cmark &
%   % \xmark &
%   \makecell[l]{Across domain; \\ Revision history} \\ % \hline
% Improving Iterative Text Revision by Learning   Where to Edit from Other Revision Tasks &
\citet{kim-etal-2022-improving}$^{\blacklozenge}$ &
  P, S  &
    367K &
  \cmark &
  \cmark &
  \cmark &
  \makecell[l]{ArXiv, Wikipedia\\ WikiNews} &
  % \xmark &
  \makecell[l]{Across domain, Revision history, \\Including edit   tagging} \\ \hline

  % Modeling the Relationship between User Comments   and Edits in Doc. Revision &
\citet{zhang-etal-2019-modeling} &
  Sec &
  786K &
  \xmark &
  \xmark &
  \cmark &
  Wikipedia &
  % \xmark &
  \makecell[l]{Revision history, Comment} \\ \hline
  % One Doc., Many Revisions: A Dataset for   Classification and Description of Edit Intents &
\citet{rajagopal-etal-2022-one} &
  D &
  9.3K &
  \cmark &
  \cmark &
  \xmark &
  Wikipedia &
  % \xmark &
  \makecell[l]{Revision history, Comment} \\ \hline
  % SWIPE: A Dataset for Doc.-Level   Simplification of Wikipedia Pages &
\citet{laban-etal-2023-swipe} &
  D &
  145K$^{*}$ &
  \cmark &
  \cmark &
  \cmark &
  Wikipedia &
  % \xmark &
  / \\ \hline
    % Verba Volant, Scripta Volant: Understanding Post-publication Title Changes in News Outlets & 
  % \citet{Guo2022} &
  % S &	
  % 41.9K &	
  % \xmark &	
  % US News agencis &
  % \xmark &	
  % \xmark & 
  % Revision history \\ \hline
% NewsEdits: A News Article Revision Dataset and a   Doc.-Level Reasoning Challenge &
\citet{spangher-etal-2022-newsedits} &
  D &
  40M &
  \xmark &
  \xmark &
  \cmark &
  News &
  % \xmark &
  Multilingual \\ \hline
% ARXIVEDITS: Understanding the Human Revision   Process in Scientific Writing &
\citet{jiang-etal-2022-arxivedits} &
  D, S  &
  13K &
  \cmark &
  \cmark &
  \cmark &
  ArXiv &
  % \cmark &
  / \\ \hline
  % Towards Automated Document Revision: Grammatical Error Correction, Fluency Edits, and Beyond &
\citet{mita-etal-2024-towards} &
  D, P &	
  64$^{*}$ &	
  \cmark & 
  \cmark &	
  \xmark	& 
  ACL anthology &	
  % \xmark &
  Comment \\ \hline

% Re3: A Holistic Framework and Dataset for   Modeling Collaborative Doc. Revision &
\citet{ruan-etal-2024-re3} &
  \makecell[l]{Sec, P, \\S, SS} &
  11.6K &
  \cmark &
  \cmark &
  \cmark &
  F1000Research &
  % \xmark &
  \makecell[l]{Review\&response included} 
  
  \\ \Xhline{1pt}
\end{tabular}
% \vspace{-7pt}
\caption{
Text-revision datasets. \textbf{Gran.} denotes revision granularity (S: sentence, SS: subsentence, P: paragraph, D: document, Sec: section); \textbf{\#Pairs} is the number of revised sentence pairs; \textbf{H.} indicates human annotation; \textbf{I.}/\textbf{A.} indicate edit intention/action labels; \textbf{Source} gives corpus origin; and \textbf{Features} lists additional information. $\ast$ marks revised document pairs, and $^\blacklozenge$ denotes the most recent corpus in a series. See Appendix \ref{sec:datasets_all} for the complete table.}
%\caption{The publicly released datasets of text revision. \textbf{Gran.} indicates revision granularity, including S (Sentence), SS (Subsentence), P (Paragraph), D (Document), and Sec (Section). \textbf{\# pairs} reports the number of revised sentence pairs in the dataset. \textbf{Human} indicates whether the dataset contains human-annotated labels. \textbf{Source} indicates where the corpus comes from. \textbf{Intent} and \textbf{Action} indicate whether the dataset is annotated with edit intention and edit action. \textbf{Features} shows the other information or features included in the dataset. Note: (1) * indicates the number of revised document pairs; (2) the corpora of the papers in a group (no delimiting line between them) are reused and expanded in the follow-up work. }
\label{tab:datasets}
% \vspace{-10pt}
\end{table*}

\subsection{Edit Granularity}

\paragraph{Sentence-level Revision}
Early work focuses on sentence-level revisions, constructing sentence-pair corpora, primarily from Wikipedia, to distinguish meaning-preserving and meaning-changing edits \citep{daxenberger-gurevych-2012-corpus, daxenberger-gurevych-2013-automatically}. Subsequent studies move beyond surface operations to annotate semantic edit intentions, capturing the motivations behind revisions \citep{yang-etal-2017-identifying-semantic}, including atomic insertion edits collected across multiple languages \citep{faruqui-etal-2018-wikiatomicedits}. Sentence-level datasets are also widely used in educational contexts, where student essay revisions support fine-grained analysis of writing development, with annotations ranging from per-edit multi-label intentions to single labels per sentence pair \citep{zhang-etal-2017-corpus, Kashefi2022-xs}.

\paragraph{Higher-level Revision}
While sentence-level analysis captures fine-grained edits, it can obscure broader revision patterns in extensively edited documents \cite{10.1007/978-3-319-07221-0_105}. Paragraph- and document-level datasets therefore enable more interpretable analysis of structural revision strategies, such as sentence reordering and content reorganization \cite{zhang-litman-2014-sentence, spangher-etal-2022-newsedits}. Existing higher-level revision corpora span Wikipedia, scientific, and news domains, preserving paragraph or document structure with annotations for edit actions and, in some cases, edit intentions  \citep{du-etal-2022-understanding-iterative, du-etal-2022-read, kim-etal-2022-improving, zhang-etal-2019-modeling, rajagopal-etal-2022-one}. More recent resources support multi-granularity or full-document analysis, distinguishing sentence-level intentions from document-level actions and enabling holistic modeling of complex revision workflows \citep{jiang-etal-2022-arxivedits, mita-etal-2024-towards}.

\subsection{Ground-truth Labels}

Most revision datasets rely on human annotation, which provides reliable supervision for model training and evaluation. In contrast, some datasets forego manual edit intention annotation in favor of automated labeling of edit actions, enabling large-scale analysis at reduced annotation cost \citep{zhang-etal-2019-modeling, spangher-etal-2022-newsedits}.

\subsection{Feature-augmented Datasets}

\paragraph{Revision History}
Many datasets preserve full revision histories, providing longitudinal signals for studying document evolution and iterative writing behavior \citep{zhang-etal-2017-corpus, Kashefi2022-xs}. Such histories enable analysis of how edit intentions and actions change over time and support modeling of iterative revision processes across domains \citep{du-etal-2022-understanding-iterative, du-etal-2022-read, kim-etal-2022-improving}.

% Many revision corpora preserve full revision histories, providing valuable signals for studying document evolution and iterative writing behavior. Datasets derived from student essays include multiple drafts capturing successive rounds of revision, enabling analysis of writing development over time \citep{zhang-etal-2017-corpus, Kashefi2022-xs}. Similarly, several large-scale corpora contain iteratively revised texts across multiple domains, supporting research on automatic iterative revision systems and longitudinal editing behavior \citep{du-etal-2022-understanding-iterative, du-etal-2022-read, kim-etal-2022-improving}.

\paragraph{Comments}
Revision comments provide explicit signals of editors’ intentions and are therefore valuable supervision for modeling revision behavior and downstream tasks. Several datasets pair edits with corresponding comments and associated metadata, enabling discriminative modeling and richer analysis of editing practices \citep{zhang-etal-2019-modeling, rajagopal-etal-2022-one, mita-etal-2024-towards}.

% Revision comments explicitly articulate an editor’s intent and therefore constitute a rich source of supervision for modeling revision behavior and downstream applications such as comment generation. Some datasets pair edits with corresponding comments and include non-matching edits or comments as negative examples, facilitating discriminative modeling \citep{zhang-etal-2019-modeling}. Other resources provide free-form textual descriptions accompanying edits, together with rich metadata such as timestamps, authorship, URLs, and revision type, enabling more detailed analysis of editing practices \citep{rajagopal-etal-2022-one}. In academic writing contexts, revisions and comments have also been produced by professional editors with extensive experience in English scholarly editing, offering high-quality annotations grounded in expert practice \citep{mita-etal-2024-towards}.

\begin{figure*}[t]
    \centering
    \includegraphics[width=1\linewidth]{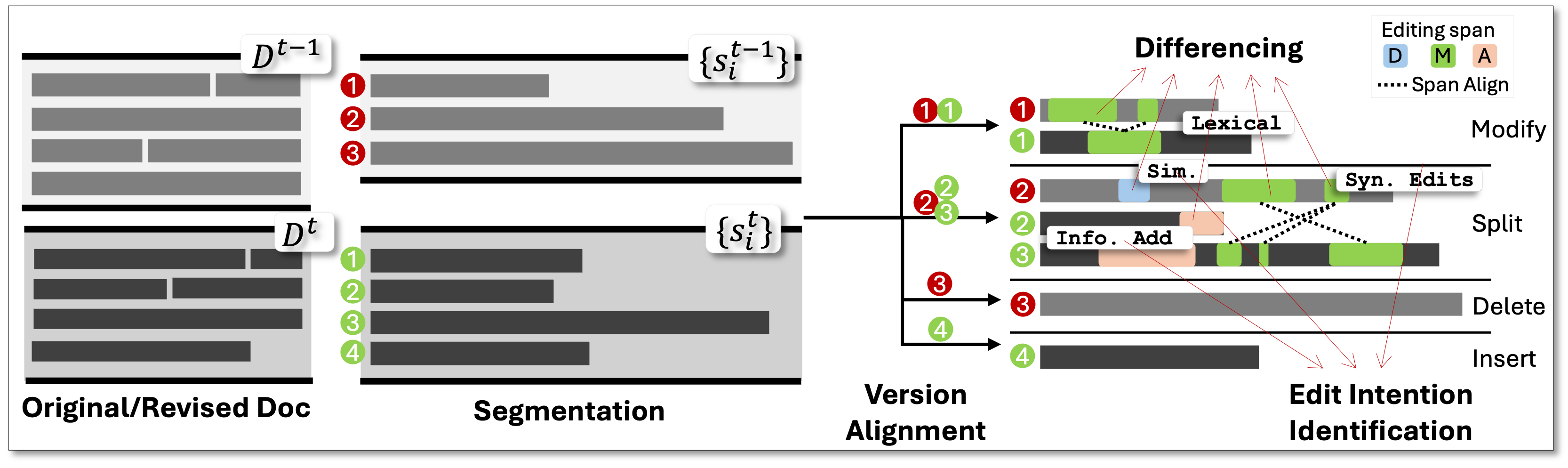}
    % \vspace{-25pt}
    \caption{The workflow of revision dataset construction. See Appendix \ref{sec:example_workflow} for a more detailed example.}
    \label{fig:edit_workflow}
    % \vspace{-10pt}
\end{figure*}

\paragraph{Reviews and Responses}
Collaborative text production typically follows cycles of drafting, peer review, revision, and response, where response documents explicitly describe the changes made in reaction to reviewer feedback \citep{cheng-etal-2020-ape, 10.1162/coli_a_00455}. Recently introduced datasets model this complete collaborative cycle, providing annotations for both edit actions and edit intentions in the context of scholarly publishing and peer review \citep{ruan-etal-2024-re3}.
\section{Revision Dataset Construction}
\label{sec:dataset_construction}

Figure~\ref{fig:edit_workflow} illustrates the core steps of revision dataset construction: segmentation, version alignment, and differencing. We briefly summarize each step and highlight the key design choices that affect downstream edit intention analysis.

\paragraph{Segmentation}
Segmentation partitions documents into revision units at a chosen granularity $g\in\{P, S\}$, enabling alignment, differencing, and annotation. Finer-grained segmentation supports associating edits with coherent intentions and improves annotation reliability, while coarser segmentation can conflate heterogeneous revising goals. Existing approaches broadly fall into rule-based methods using punctuation heuristics \cite{lee-webster-2012-corpus} and learning-based models that treat boundary detection as a structured prediction problem \citep{ reynar-ratnaparkhi-1997-maximum}.

\paragraph{Version Alignment}
Version alignment establishes correspondences between texts at the same granularity across versions, $g^t-1$ and $g^t$, identifying unchanged, revised, deleted, and added units. Most alignment methods rely on similarity estimation followed by a decision procedure. Similarity can be computed using surface-level lexical measures or semantic representations derived from language models, with the latter better handling paraphrasing and contextual reformulation \citep{zhang-litman-2014-sentence, jiang-etal-2020-neural}. To improve scalability and consistency, prior work adopts hierarchical alignment (e.g., paragraph-to-sentence) or joint structured prediction to avoid exhaustive pairwise matching \citep{jiang-etal-2020-neural, ruan-etal-2024-re3}.

\paragraph{Differencing}
Differencing identifies the revised text spans ($\{p_{k}\}$) between aligned texts. Classical approaches apply string-matching-based algorithms such as longest common subsequence or edit distance to extract insertions, deletions, and substitutions \citep{Myers1986-qu}. While efficient, these methods operate purely at the surface level and often conflate semantically distinct edit actions or fail to capture paraphrastic substitutions and text movement. Recent approaches reformulate differencing as semantic span alignment, producing finer-grained and more interpretable edit units that better support edit intention identification \citep{jiang-etal-2022-arxivedits}. See Appendix \ref{sec:demo_diff_alg} for an example.

Overall, revision corpus construction requires coordinated design choices across segmentation, alignment, and differencing, as errors or mismatches at early stages directly propagate to edit intention annotation and modeling.

\section{Edit Intention Taxonomy}
\label{sec:eit_intention_categorization}

%begin{comment}
% We now turn to how the motivations underlying revisions are conceptualized and organized through EITs. 
% Edit intention categorization refers to the process of identifying the full range of intentions underlying textual revisions and organizing them into an EIT. 
% In this section, we focus on the construction principles and design choices behind EITs, rather than enumerating specific edit intention labels. 
We now turn to how the motivations underlying revisions are conceptualized and organized.
Edit intention categorization refers to the process of identifying the full range of intentions underlying textual revisions and organizing them into an EIT.
Beyond providing a vocabulary of revision purposes, the design of an EIT also shapes how edit intention identification is formulated (e.g., label granularity and hierarchy), how annotations are collected, and how results are compared across datasets.
In this section, we focus on the construction principles and design choices behind EITs, rather than enumerating specific edit intention labels.

% Consider the following simplified EIT commonly used in revision studies. Revisions are first distinguished by whether they preserve or change meaning. \textit{Surface-level edits} include \textsc{Fluency} (grammar and spelling correction) and \textsc{Formatting}. \textit{Meaning-changing edits} include \textsc{Elaboration} (adding new information), \textsc{Verification} (correcting factual errors or adding citations), and \textsc{Simplification} (removing redundancy or reducing complexity). This structure reflects a widely adopted taxonomy lineage in academic writing and Wikipedia revision analysis.

Consider the following simplified EIT commonly used in revision studies.
Revisions are first distinguished by whether they preserve or change meaning.
\textit{Surface-level edits} include \textsc{Fluency} (grammar and spelling correction) and \textsc{Formatting}.
\textit{Meaning-changing edits} include \textsc{Elaboration} (adding new information), \textsc{Verification} (correcting factual errors or adding citations), and \textsc{Simplification} (removing redundancy or reducing complexity).
This structure reflects a widely adopted taxonomy lineage in academic writing and Wikipedia revision analysis.

% \subsection{Statistics of Edit Intention Taxonomies}
\subsection{Characterizing Edit Intention Taxonomies}

EITs characterize the purposes behind text revisions.
In this survey, we review prior EITs along four dimensions: taxonomy structure, category definitions and revision examples, application domains, and availability of data and resources (see the last one in Appendix \ref{sec:edit_domains}; Table~\ref{tab:paper-list} summarizes the surveyed taxonomies in Appendix \ref{sec:eit_stats}).

% EITs have been widely studied as a means of characterizing the purposes behind text revisions.
% A well-designed EIT typically provides a structured set of categories with clear names, concise definitions, and representative revision examples.
% In this survey, we review prior EITs along four dimensions: structure, category definitions and revision examples, application domains, and availability of data and resources (See Appendix \ref{sec:edit_domains}).
% Table~\ref{tab:paper-list} summarizes the properties of the surveyed EITs.

% \begin{figure*}
%     \centering
%     \includegraphics[width=1\linewidth]{figures/tree3.png}
%     \vspace{-25pt}
%     \caption{Lineage of existing EITs. Each block (labeled by paper alias; see Table \ref{tab:paper-list}) represents an EIT, with dashed arrows indicating inheritance relationships. Link labels denote how child EITs derive from parent EITs, and blocks sharing the same color indicate highly similar taxonomies (e.g., Pfe06 and Liu11)}
%     \label{fig:lineage}
%     \vspace{-10pt}
% \end{figure*}

\begin{figure*}
    \centering
    \includegraphics[width=1\linewidth]{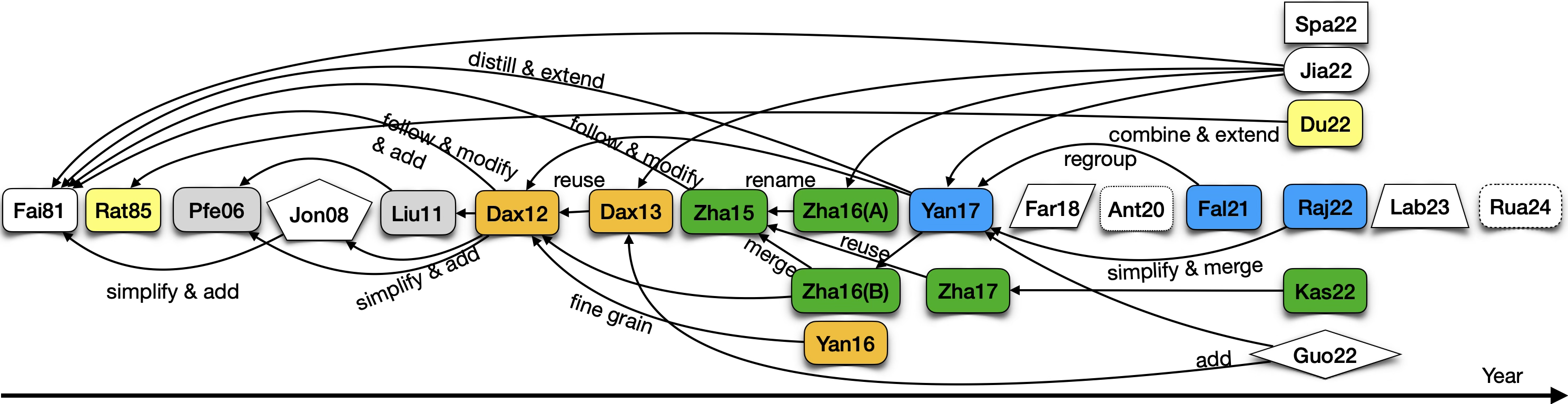}
    % \vspace{-20pt}
    \caption{Lineage of existing EITs. Each block (labeled by paper alias; see Table \ref{tab:paper-list}) represents an EIT, with dashed arrows indicating inheritance relationships. Link labels denote how child EITs derive from parent EITs, and blocks sharing the same color indicate highly similar taxonomies (e.g., Pfe06 and Liu11)}
    \label{fig:lineage}
    % \vspace{-10pt}
\end{figure*}

\paragraph{Taxonomy Structure}
Most EITs adopt a tree-based organization, ranging from flat category lists to hierarchical structures with up to three levels.
Hierarchical EITs commonly distinguish high-level revision purposes from lower-level edit actions.
Three-level taxonomies typically separate meaning-preserving from meaning-changing revisions
\cite{Faigley1981, zhang-litman-2015-annotation, Yang_Halfaker_Kraut_Hovy_2016},
sometimes further refining meaning changes based on discourse scope or edited objects.
Two-level taxonomies retain similar high-level distinctions but often omit explicit modeling of edit operations
\cite{Jones2008, daxenberger-gurevych-2012-corpus, du-etal-2022-understanding-iterative, ruan-etal-2024-re3}.
Flat taxonomies either collapse earlier hierarchies
\cite{zhang-litman-2016-using, rajagopal-etal-2022-one}
or focus on specialized edit phenomena, such as edit actions
\cite{spangher-etal-2022-newsedits},
clarity \cite{Rathjens1985},
or headline revisions \cite{Guo2022}.
Despite structural differences, many EITs capture overlapping revision concepts.

\paragraph{Definitions and Revision Examples}
Clear definitions and representative examples are critical for EIT usability \cite{TheRoleofTaxonomy}.
Omissions are common, particularly when category names appear self-explanatory
\cite{anthonio-etal-2020-wikihowtoimprove},
and some categories lack both definitions and examples, leading to ambiguity (e.g., \emph{Style and Readability} \cite{Jones2008}).
Definition quality also varies, with some categories being overly broad (e.g., \emph{Copy-Editing}) or inconsistent with their descriptions.
Examples are typically provided only for leaf categories, and coverage is uneven across taxonomies
\cite{Faigley1981, Jones2008, Liu2011, faltings-etal-2021-text},
complicating interpretation and cross-taxonomy comparison.

% \paragraph{Key Takeaways}
% Across existing EITs, design choices vary substantially in structure and granularity, and the quality of definitions/examples is uneven.
% These variations affect both annotation usability and the comparability of edit intention labels across datasets, motivating a closer look at how EITs are constructed and evolve.

\paragraph{Application Domains}
EITs have been developed across diverse application domains.
A substantial portion of the literature focuses on Wikipedia, examining collaborative behaviors, editor roles, and semantic edit intentions
\cite{Pfeil2006, Jones2008, daxenberger-gurevych-2012-corpus, Yang_Halfaker_Kraut_Hovy_2016, yang-etal-2017-identifying-semantic}.
Writing-focused studies analyze revisions in student, technical, and academic texts to understand revision intent, clarity, and argumentative development
\cite{Faigley1981, Rathjens1985, zhang-litman-2015-annotation, Kashefi2022-xs}.
Other work targets news articles and headlines
\cite{spangher-etal-2022-newsedits, Guo2022}
or instructional texts such as wikiHow
\cite{anthonio-etal-2020-wikihowtoimprove}.
More recent EITs aim to generalize across multiple domains
\cite{du-etal-2022-understanding-iterative}.

Overall, existing EITs exhibit substantial conceptual overlap but differ in structure, definition quality, domain coverage, and resource support.

\subsection{EIT Construction}
We analyze relationships among EITs through their citation network (Figure~\ref{fig:lineage}; see Appendix \ref{sec:lineage} for details), which reveals an EIT lineage:
later EITs often build on earlier ones via merging, splitting, and renaming categories.
% We use this lineage-aware view in the next section to group EITs for group-wise integration.

Constructing an EIT is non-trivial because EITs evolve with task objectives, application domains, and annotation requirements.
For example, the EIT proposed for academic writing \citep{zhang-litman-2015-annotation} is later adapted to analyze semantic edit intentions in Wikipedia revisions \citep{yang-etal-2017-identifying-semantic}.
A follow-up study collapses all subcategories under \textit{Surface Changes}, reflecting a shift toward augmentative revisions \citep{zhang-litman-2016-using}.

A lineage-aware perspective shows that many EITs evolve via systematic transformations rather than purely bottom-up or top-down design \cite{lan-etal-2025-unit}.
For instance, synthesizing prior EITs and iteratively refining them through batch annotation of samples enables the emergence of document-level categories not observable in sentence-level revision analysis \citep{laban-etal-2023-swipe}.
Common evolution patterns include:
(1) \textit{reuse-based pattern};
(2) \textit{merge-based pattern};
(3) \textit{refinement-based pattern}; and
(4) \textit{hierarchy adaptation}.
These patterns highlight that bottom-up and top-down strategies are not mutually exclusive.

\section{Methods: Edit Intention Identification}
\label{sec:edit_intention_identification}

% 1. Human annotation: cost / quality / scalability; IAA
% 2. Supervised & neural models: feature vs representation; do not mention 
% 3. LLM-based approaches: ICL、prompt sensitivity、reproducibility

%\subsection{Edit Intention Annotation}

For the edits in a revision, $\{e_{k}\} = R^{t,g}$, we may annotate one or more edit intentions from an EIT, $\{I_{k}\}\in EIT$, that most capture the editor's intention. Due to the semantic complexity and implicit nature of edit intentions, most existing studies rely on manual annotation to construct edit intention taxonomies, create revision datasets with gold labels, and analyze revision behavior, despite the substantial cost and limited scalability of this approach \citep{yang-etal-2017-identifying-semantic, Kashefi2022-xs}.

\paragraph{Manual Annotation}
Manual annotation is typically conducted via crowdsourcing platforms such as Amazon Mechanical Turk or through trained students, with annotator selection emphasizing language proficiency and, in some cases, domain or editing expertise \citep{zhang-etal-2017-corpus, daxenberger-gurevych-2012-corpus}. For example, Wikipedia-based corpora often require familiarity with platform conventions and policies \citep{yang-etal-2017-identifying-semantic}. To ensure annotation consistency, studies commonly employ structured training procedures that include detailed guidelines, illustrative examples, live demonstrations, and iterative practice with feedback \citep{yang-etal-2017-identifying-semantic, du-etal-2022-understanding-iterative, lan-etal-2025-unit}. Annotation reliability is assessed using inter-annotator agreement metrics, with Krippendorff’s $\alpha$ and MASI frequently adopted in multi-label or overlapping intention settings \citep{daxenberger-gurevych-2012-corpus, passonneau-2006-measuring}.

\paragraph{Automatic Annotation}
Automatic edit intention identification has been explored using both traditional machine learning and neural approaches. Early work formulates the task as binary or multi-label classification using feature-based models, often employing one-vs-rest strategies and designed features derived from textual differences, discourse cues, and metadata \citep{zhang-litman-2015-annotation, daxenberger-gurevych-2013-automatically}. While interpretable and computationally efficient, such models struggle with paraphrasing and context-dependent semantic changes. More recent approaches adopt neural and LLMs that encode original and revised text jointly and better capture semantic intent \citep{jiang-etal-2022-arxivedits}. In particular, in-context learning with LLMs enables flexible intent generation without task-specific training, but introduces challenges related to computational cost, reproducibility, and output consistency \citep{ruan-etal-2024-re3}.  Appendix \ref{sec:eval_metrics} gives the annotation evaluation metrics.

\paragraph{Trade-offs}
Manual and automatic annotation methods exhibit complementary strengths and limitations. Manual annotation provides high-fidelity data grounded in human judgment and is essential for EIT development and further analysis, but is expensive and hard to scale \citep{yang-etal-2017-identifying-semantic}. Automatic approaches enable large-scale analysis across domains and revision histories, but remain sensitive to domain shift, annotation guidelines, and model assumptions \citep{zhang-litman-2015-annotation}. Thus, many studies adopt hybrid strategies in which manually annotated datasets serve as gold standards for training, evaluation, or calibration of automatic models, balancing annotation quality with scalability \citep{jiang-etal-2022-arxivedits, ruan-etal-2024-re3}.

\section{Application}
\label{sec:application}

Edit intention has been adopted as a unifying abstraction across a range of applications that seek to analyze, support, or summarize revision behavior beyond surface-level textual diffs.

\paragraph{Revision Behavior Analysis}
Edit intentions have been widely used to study collaboration dynamics in online collaboration systems. Prior work shows that different intention distributions correlate with editor retention, participation patterns, and quality outcomes \citep{yang-etal-2017-identifying-semantic}. For example, maintenance- and integration-oriented edits are associated with long-term editor survival, whereas simplification- or vandalism-related edits correlate with higher revert rates and early disengagement. Edit intentions have also been shown to contribute differently across an article’s lifecycle \cite{yang-etal-2017-identifying-semantic}, with content-expanding intentions playing a larger role in early stages and refinement-oriented intentions becoming more prominent as articles mature.

Beyond individual behavior, edit intentions provide a lens for analyzing sociocultural variation and division of labor in collaborative writing. Cross-lingual studies reveal differences in revision strategies across cultural contexts, indicating that revision behavior is shaped by social norms and task demands \citep{Pfeil2006}. Other work models editor roles by aggregating intention distributions over revision histories, uncovering interpretable mixtures of roles such as substantive contributors, copy editors, and vandal fighters, and demonstrating how different roles contribute to quality over time \citep{Yang_Halfaker_Kraut_Hovy_2016, Liu2011}.

\paragraph{Writing Assistance Systems}
Edit intention has also been used as a control signal in writing assistance systems. Intention-aware editors model revision as a goal-directed process, enabling users to request specific types of changes (e.g., improving fluency versus adding content) and supporting iterative, interactive revision workflows. Empirical results show that conditioning generation on explicit edit intentions leads to more effective revisions than intent-agnostic baselines \citep{faltings-etal-2021-text}. More approaches combine edit intention with free-form edit descriptions, using intention as a coarse semantic scaffold and natural language explanations to capture finer-grained rationale. This hybrid representation improves revision generation, reinforcing edit intention as a foundational abstraction for controllable and interpretable writing support \citep{rajagopal-etal-2022-one}.

\paragraph{Document Edit Summarization}
Document edit summarization represents a higher-level application of edit intention analysis, aiming to abstract low-level diffs into concise summaries that explain what changed and why. Early work aggregates categorized edits to produce structured summaries of revision activity \cite{10.1145/1832772.1832775}, while more recent formulations treat summarization as an intent-aware generation task  \citep{ruan-etal-2024-re3}. By leveraging edit actions, intentions, and document structure, models generate natural-language summaries that resemble human revision descriptions. Results show that edit intention provides essential semantic guidance for summarization, though challenges remain in coverage, factual grounding, and discourse coherence for large documents.

\paragraph{Additional Uses}
Edit intention has also been applied in educational writing assessment, where revision types are used to characterize student writing development and support targeted feedback \citep{Sommers1980, FaigleyWitte1981, zhang-etal-2017-corpus, Kashefi2022-xs}. In collaborative platforms, certain intentions (e.g., vandalism or revert-triggering edits) serve as signals for moderation, quality control, and editor modeling \citep{Potthast2008, Halfaker2013}. More recently, edit intention has been explored as a semantic abstraction for evaluating and interpreting model-generated revisions, enabling intent-aware comparison between human and automated editing behavior \citep{jiang-etal-2022-arxivedits, ruan-etal-2024-re3}.

\section{Future Research Directions}
\label{sec:limitations}

%\subsection{Future Research Directions}

\paragraph{Evaluation and Benchmark Design} Progress in edit intention research would benefit from evaluation frameworks that balance standardization with flexibility. Rather than enforcing a single universal taxonomy, future benchmarks could define shared core intention sets alongside domain-specific extensions, enabling both comparability and expressiveness. Designing evaluation protocols that account for overlapping or multi-intention edits remains an important research direction.

\paragraph{Evolving and Generalizable Edit Intention Taxonomies}
Future work may move toward edit intention taxonomies that are explicitly designed to evolve across domains, tasks, and revision granularities. Rather than treating taxonomies as fixed label inventories, research may emphasize extensible designs that support refinement, aggregation, and specialization while preserving core conceptual distinctions. Lineage-aware taxonomy construction, where new categories are introduced through principled reuse, merging, or refinement of existing ones, offers a path toward improving comparability and cumulative progress across studies.

%\paragraph{Robust and Interpretable Edit Intention Identification} Advancing edit intention identification requires models that are robust across domains, transparent in their predictions, and capable of handling fine-grained and multi-intention edits. While LLMs have shown encouraging results in capturing revision semantics, future research may focus on improving controllability, explainability, and alignment with human annotation standards. Hybrid approaches that combine symbolic structure, learned representations, and human feedback may help bridge the gap between research prototypes and deployable systems.

\paragraph{Multi-intention and Cross-granularity Modeling}
Most existing work assigns a single intention to each edit at a fixed granularity, typically the sentence level. Future research should develop models that support multi-intention labeling and explicitly link sentence-level edits to paragraph- and document-level revising goals, such as restructuring, argument development, or narrative flow. Such representations would better reflect real-world revision behavior.

\paragraph{Process-level Modeling of Revision Dynamics}
Revisions are inherently sequential and iterative, yet are often modeled as independent edits in isolation. Capturing how edit intentions evolve over time, and how earlier revisions constrain or enable later ones, offers a promising direction for understanding revision strategies and improving downstream applications such as writing assistance and revision planning.

\paragraph{Future Applications}
Edit intention modeling has demonstrated value in revision behavior analysis, writing assistance, and document edit summarization, and continues to open new application opportunities. 
Promising directions include intent-aware software documentation maintenance, incremental summary updating, educational feedback and writing assessment, moderation and quality control in collaborative platforms, and evaluation of model-generated revisions.
In high-stakes domains such as legal, medical, or policy drafting, edit intentions may further support auditability, justification, and explainability of revisions. While many of them remain exploratory, they underscore edit intention as a unifying abstraction for understanding, guiding, and evaluating text revision across domains.
Edit intentions also offer a principled lens to study LLM evolution by systematically characterizing how factual knowledge, temporal awareness, and writing behavior change across model versions through analyzing intentional differences in responses to the same prompts over time.

\section{Conclusion}

This survey synthesizes text revision research through the unified lens of \emph{edit intentions}, organizing prior work across datasets, corpus construction workflows, EIT design, identification methods, and downstream applications. We further introduce a structured categorization framework and highlight how EITs evolve over time and across domains through a citation network of existing EITs, helping capture EIT construction principles and design choices in the literature. 
By reviewing both manual and automatic annotation paradigms and mapping them to downstream applications, we show how edit intention serves as a practical semantic abstraction beyond surface diffs. Finally, we outline key challenges and dedicate future research to exploring this area in promising directions.  

% This survey synthesizes text revision research through the unified lens of \emph{edit intentions}, organizing prior work across datasets, corpus construction workflows, EIT design, identification methods, and downstream applications.

% We further introduce a structured categorization framework and highlight how EITs evolve across domains via lineage-aware operations such as merging, splitting, and refinement, helping reconcile fragmented terminology and design choices in the literature. 
% By reviewing both manual and automatic identification paradigms and mapping them to applications including revision behavior analysis, writing assistance, and document edit summarization, we show how edit intention serves as a practical semantic abstraction beyond surface diffs.
% Finally, we outline key challenges—taxonomy inconsistency, scalability and robustness of identification, and lack of standardized evaluation—and point to promising directions such as richer multi-intention representations, cross-granularity modeling, scalable human–model annotation, and process-level revision modeling. 

\section{Acknowledgements}
This work was supported in part by the U.S. National Science Foundation awards III-2107213, 2026513, and ITE-2333789.

\section{Limitations}

This survey has several limitations that reflect both the scope of the paper and the current state of edit intention research.

First, our coverage is necessarily bounded by the availability of publicly released datasets and published work. Although we made a best effort to include representative studies across domains and revision granularities, relevant datasets or methods may have been overlooked, particularly in non-English settings or in proprietary writing systems that do not release revision histories.

Second, the diversity of edit intention taxonomies poses an inherent challenge for comprehensive synthesis. Edit intentions are often domain-dependent, overlapping, and evolving, which limits direct comparability across studies. While we emphasize lineage-aware analysis and highlight common construction patterns, our categorization cannot fully reconcile differences in label definitions or annotation guidelines across datasets.

Finally, recent advances in large language models are evolving rapidly, and some findings summarized in this survey may be affected by future model releases or methodological shifts. Although we aim to capture stable research directions and enduring challenges, specific modeling approaches or empirical conclusions may require re-evaluation as the field progresses.

%%%%%%%%%%%%%%%% ACL2026 version end %%%%%%%%%%%%%%%

% \begin{acks}
% \end{acks}

% \bibliographystyle{ACM-Reference-Format}
\bibliography{taxonomy, metrices, schema_matching, db_integration, lab_papers}

\appendix

% %%%%%%%%%%%%%%% Long paper version begin %%%%%%%%%%%%%%%
% \input{sections/appendix}
% %%%%%%%%%%%%%%%% Long paper version end %%%%%%%%%%%%%%%

%%%%%%%%%%%%%%% ACL2026 version begin %%%%%%%%%%%%%%%

\section{Complete Research Taxonomy}
\label{sec:complete_research_taxonomy}

We present the complete taxonomy of research in text revision in the lens of edit intention in Figure \ref{fig:taxonomy-all}.

\tikzstyle{my-box}=[
    rectangle,
    draw=gray,
    rounded corners,
    text opacity=1,
    minimum height=1.5em,
    minimum width=5em,
    inner sep=2pt,
    align=center,
    fill opacity=.5,
    line width=0.8pt,
]
\tikzstyle{leaf}=[my-box, minimum height=1.5em,
    fill=pink!10, text=black, align=left,font=\normalsize,
    inner xsep=2pt,
    inner ysep=4pt,
    line width=0.8pt,
]

\definecolor{c1}{RGB}{93,191,237} % blue
\definecolor{c2}{RGB}{237,110,106} % red
\definecolor{c3}{RGB}{240,154,69} % yellow
\definecolor{c4}{RGB}{8,153,68} % green
\definecolor{c5}{RGB}{205,180,243} % purple
\definecolor{c6}{RGB}{97,218,184} % cyan
\definecolor{c7}{RGB}{0,128,128} % teal

\begin{figure*}[!t]
    \centering
    \resizebox{\textwidth}{!}{
        \begin{forest}
            forked edges,
            for tree={
                grow=east,
                reversed=true,
                anchor=base west,
                parent anchor=east,
                child anchor=west,
                base=center,
                font=\large,
                rectangle,
                % draw=hidden-draw,
                draw=gray,
                rounded corners,
                align=left,
                text centered,
                minimum width=4em,
                edge+={darkgray, line width=1pt},
                s sep=3pt,
                inner xsep=2pt,
                inner ysep=3pt,
                line width=0.8pt,
                ver/.style={rotate=90, child anchor=north, parent anchor=south, anchor=center},
            },
            where level=1{text width=12em,font=\normalsize,}{},
            where level=2{text width=12em,font=\normalsize,}{},
            where level=3{text width=10em,font=\normalsize,}{},
            where level=4{text width=35em,font=\normalsize,}{},
            where level=5{text width=18em,font=\normalsize,}{},
            [
                \textbf{Edit/Revision Intentions}, ver, line width=0.7mm
                [
                    \textbf{Datasets}, fill=c1!60, draw=c1, line width=0mm
                    [   
                        \textbf{Edit Granularity}, fill=c1!60, draw=c1, line width=0mm, edge={c1}
                        [
                            Sentence-level Revision: \citet{zhang-etal-2017-corpus, Kashefi2022-xs}; \\
                            $\ \ $ \citet{faruqui-etal-2018-wikiatomicedits, anthonio-etal-2020-wikihowtoimprove, bhat-etal-2020-towards}; \\
                            Document-level Revision: \citet{spangher-etal-2022-newsedits, laban-etal-2023-swipe}; \\
                            Multi-level Revision: \citet{du-etal-2022-understanding-iterative, du-etal-2022-read, kim-etal-2022-improving}; \\ 
                            $\ \ $ \citet{jiang-etal-2022-arxivedits, Kashefi2022-xs, ruan-etal-2024-re3};
                            , leaf, text width=32em, draw=c1, line width=0.7mm, edge={c1}
                        ]
                    ]
                    [
                        \textbf{Annotated Edits}, fill=c1!60, draw=c1, line width=0mm, edge={c1}
                        [
                              Multi-labeled Edits: \citet{daxenberger-gurevych-2012-corpus, daxenberger-gurevych-2013-automatically}; \\ 
                              $\ \ $ \citet{yang-etal-2017-identifying-semantic};\\
                              Edit intention labeled: \citet{daxenberger-gurevych-2012-corpus, daxenberger-gurevych-2013-automatically}; \\
                              $\ \ $ \citet{yang-etal-2017-identifying-semantic, faruqui-etal-2018-wikiatomicedits, zhang-etal-2017-corpus};\\
                              $\ \ $ \citet{Kashefi2022-xs, du-etal-2022-understanding-iterative, du-etal-2022-read, kim-etal-2022-improving}; \\
                              $\ \ $ \citet{rajagopal-etal-2022-one, laban-etal-2023-swipe, jiang-etal-2022-arxivedits}; \\ 
                              $\ \ $ \citet{mita-etal-2024-towards, ruan-etal-2024-re3}; \\
                              Edit action labeled: \citet{daxenberger-gurevych-2012-corpus, daxenberger-gurevych-2013-automatically}; \\ 
                              $\ \ $ \citet{zhang-etal-2017-corpus, Kashefi2022-xs, du-etal-2022-understanding-iterative, du-etal-2022-read}; \\ 
                              $\ \ $ \citet{kim-etal-2022-improving, laban-etal-2023-swipe, spangher-etal-2022-newsedits}; \\ 
                              $\ \ $ \citet{ jiang-etal-2022-arxivedits, ruan-etal-2024-re3};\\
                              , leaf, text width=32em, draw=c1, line width=0.7mm, edge={c1}
                        ]
                    ]
                    [
                        \textbf{Feature-augmented Edits}, fill=c1!60, draw=c1, line width=0mm, edge={c1}
                        [
                              Layout: \citet{daxenberger-gurevych-2012-corpus, daxenberger-gurevych-2013-automatically}; \\ 
                              Comments: \citet{lee-webster-2012-corpus, zhang-etal-2019-modeling}; \\ 
                              $\ \ $ \citet{ rajagopal-etal-2022-one, ruan-etal-2024-re3};  \\
                              Review/response: \citet{ruan-etal-2024-re3}; \\
                              Multilingual: \citet{faruqui-etal-2018-wikiatomicedits, spangher-etal-2022-newsedits}; 
                              ,leaf, text width=32em, draw=c1, line width=0.7mm, edge={c1}
                        ]
                    ]
                ]% level1
                % [
                %     \textbf{Segmentation}, align=center,
                %     fill=c2!60, draw=c2, line width=0mm
                %     [
                %         Dynamic Programming; Regex Rules; CRF model; Tokenizer ,leaf, text width=32em, draw=c2, line width=0.7mm, edge={c2}
                %     ]
                % ]% level1
                [
                    \textbf{Revision Dataset } \\\textbf{Construction}, align=center, fill=c3!60, draw=c3, line width=0mm
                    [
                        \textbf{Segmentation}, fill=c3!60, draw=c3, line width=0mm, edge={c3}
                        [
                         Rule-based: \citet{lee-webster-2012-corpus, bronner-monz-2012-user}; \\
                         Learning-based: \citet{reynar-ratnaparkhi-1997-maximum}; \\
                        , leaf, text width=32em, draw=c3, line width=0.7mm, edge={c3}
                        ]
                    ]
                    [
                        \textbf{Version Alignment}, fill=c3!60, draw=c3, line width=0mm, edge={c3}
                        [
                         Lexical Similarity-Based: \citet{lee-webster-2012-corpus, jiang-etal-2022-arxivedits}; \\ 
                              $\ \ $ \citet{daxenberger-gurevych-2012-corpus, tan-lee-2014-corpus, ruan-etal-2024-re3}; \\ 
                              $\ \ $ \citet{ faruqui-etal-2018-wikiatomicedits, anthonio-etal-2020-wikihowtoimprove, zhang-etal-2017-corpus}; \\ 
                              $\ \ $ \citet{ zhang-litman-2014-sentence, zhang-litman-2015-annotation, zhang-litman-2016-using, Kashefi2022-xs}; \\ 
                         Semantic Similarity-Based: \citet{jiang-etal-2020-neural, ruan-etal-2024-re3};\\
                        , leaf, text width=32em, draw=c3, line width=0.7mm, edge={c3}
                        ]
                    ]
                    [
                        \textbf{Differencing}, fill=c3!60, draw=c3, line width=0mm, edge={c3}
                        [
                        String-matching-based: \citet{yang-etal-2017-identifying-semantic, Kashefi2022-xs};\\
                        $\ \ $ \citet{du-etal-2022-understanding-iterative, daxenberger-gurevych-2012-corpus} \\
                        Span alignment: \citet{jiang-etal-2022-arxivedits}; \\
                        Section-title based: \citet{tan-lee-2014-corpus}; \\
                        , leaf, text width=32em, draw=c3, line width=0.7mm, edge={c3}
                        ]
                    ]
                ]% level1
                [
                    \textbf{Edit Intention} \\\textbf{Taxonomy}, align=center, fill=c2!60, draw=c2, line width=0mm
                    [
                        \textbf{Structure},  fill=c2!60, draw=c2, line width=0mm, edge={c2}
                        [
                        Hierarchical: \citet{Faigley1981, zhang-litman-2015-annotation}; \\
                        Flat: \citet{zhang-etal-2016-argrewrite, rajagopal-etal-2022-one, Rathjens1985}
                        , leaf, text width=32em, draw=c2, line width=0.7mm, edge={c2}
                        ]
                    ]
                    [
                        \textbf{Application Domain},  fill=c2!60, draw=c2, line width=0mm, edge={c2}
                        [
                         Wikipedia: \citet{Pfeil2006, yang-etal-2017-identifying-semantic};\\
                         Writings: \citet{Faigley1981, Kashefi2022-xs}; \\
                         News: \citet{spangher-etal-2022-newsedits, Guo2022}; \\
                         Others: \citet{anthonio-etal-2020-wikihowtoimprove}
                        , leaf, text width=32em, draw=c2, line width=0.7mm, edge={c2}
                        ]
                    ]
                ]% level1
                [
                    \textbf{Edit Intention}\\\textbf{Identification}, align=center, fill=c5!60, draw=c5, line width=0mm
                    [ 
                        \textbf{Crowdsourcing}, fill=c5!60, draw=c5, line width=0mm, edge={c5}
                        [
                        \citet{zhang-litman-2015-annotation, zhang-etal-2016-argrewrite}
                        , leaf, text width=32em, draw=c5, line width=0.7mm, edge={c5}
                        ]
                    ]
                    [ 
                        \textbf{Automatic Annotation}, align=center, fill=c5!60, draw=c5, line width=0mm, edge={c5}
                        [
                        Statistical Machine Learning: \citet{zhang-litman-2015-annotation, zhang-etal-2016-argrewrite} 
                        \\
                        $\ \ $  \citet{yang-etal-2017-identifying-semantic, Yang_Halfaker_Kraut_Hovy_2016, zhang-etal-2017-corpus}; 
                        \\
                        $\ \ $ \citet{daxenberger-gurevych-2013-automatically}
                        \\
                        Neural Language Models: \citet{jiang-etal-2022-arxivedits, ruan-etal-2024-re3}
                        , leaf, text width=32em, draw=c5, line width=0.7mm, edge={c5}
                        ]
                    ]
                ]% level1
                [
                    \textbf{Applications}, fill=c6!60, draw=c6, line width=0mm
                    [   
                        Revision Behavior Studies: \citet{yang-etal-2017-identifying-semantic, Pfeil2006, Yang_Halfaker_Kraut_Hovy_2016, Liu2011} \\
                        Writing Assistance System: \citet{faltings-etal-2021-text, rajagopal-etal-2022-one} \\
                        Document Edit Summarization: \citet{ruan-etal-2024-re3}
                        % \textbf{Revision Behavior Studies},  fill=c6!60, draw=c6, line width=0mm, edge={c6}
                        % [
                        % Interactive Text Editing \cite{faltings-etal-2021-text}; \\
                        % Iterative Text Editing \cite{du-etal-2022-understanding-iterative}; 
                        , leaf, text width=45.5em, draw=c6, line width=0.7mm, edge={c6}
                        ]
                    ]
                ]
            ]
        \end{forest}
    }
    \vspace{-0mm}
    \caption{Taxonomy of edit intention-related research.}
    \label{fig:taxonomy-all}
    \vspace{0mm}
\end{figure*}
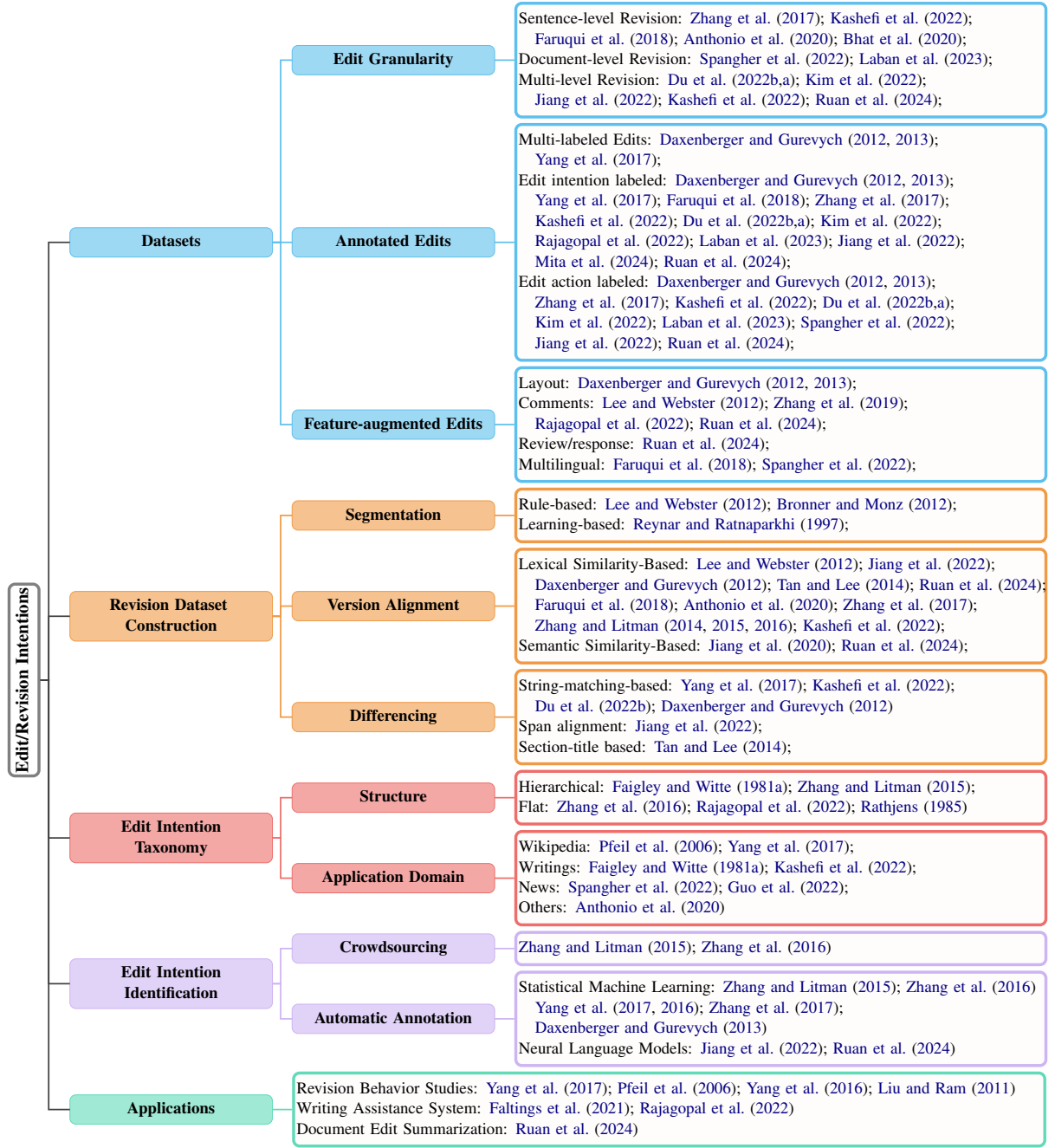

\section{Complete Revision Datasets}
\label{sec:datasets_all}

We list the complete list of the corpora of text revision with edit intentions in Table \ref{tab:datasets-all}.

\begin{table*}[t]
\centering
%\small
\footnotesize
\begin{tabular}{p{0.17\linewidth}|l|l|l|l|l|l|l}
% \hline
\Xhline{1pt}
\textbf{Paper} &
  \textbf{Gran.} &
  \textbf{\# pairs} &
  \textbf{Human} &
  \textbf{Source} &
  \textbf{Intent} &
  \textbf{Action} &
  % \textbf{Multi labels} &
  \textbf{Features} \\ \hline
  \citet{daxenberger-gurevych-2012-corpus}
% A Corpus-Based Study of Edit Categories in   Featured and Non-Featured Wikipedia Articles 
&
  S &
  2K &
  \cmark &
  Wikipedia &
  \cmark & 
  \cmark &
  % \cmark &
  Including   layout \\ % \hline
  
% Automatically Classifying Edit Categories in   Wikipedia Revisions &
\citet{daxenberger-gurevych-2013-automatically} &
  S &
  2K &
  \cmark &
  Wikipedia &
  \cmark &
  \cmark &
  % \cmark &
  Including   layout \\ \hline
  % Identifying Semantic Edit Intentions from   Revisions in Wikipedia &
\citet{yang-etal-2017-identifying-semantic} &
  S &
  5.7K &
  \cmark &
  Wikipedia &
  \cmark &
  \xmark &
  % \cmark &
  Multiple   labels \\ \hline
  
% WikiAtomicEdits: A Multilingual Corpus of   Wikipedia Edits for Modeling Language and Discourse &
\citet{faruqui-etal-2018-wikiatomicedits} &
  S &
  43M &
  \cmark &
  Wikipedia &
  \cmark &
  \xmark &
  % \xmark &
  Multilingual \\ \hline

% A   Corpus of Annotated Revisions for Studying Argumentative Writing &
\citet{zhang-etal-2017-corpus} &
  S &
  180 &
  \cmark &
  Student   Essays &
  \cmark &
  \cmark &
  % \xmark  &
  Revision   history \\ % \hline
  % ArgRewrite V.2: an Annotated Argumentative   Revisions Corpus &
\citet{Kashefi2022-xs} &
  S &
  3K &
  \cmark &
  Student   Essays &
  \cmark &
  \cmark &
  % \xmark &
  \makecell[l]{Revision   history; \\ Multiple labels} \\ \hline

% % wikiHowToImprove: A Resource and Analyses on   Edits in Instructional Texts &
% \citet{anthonio-etal-2020-wikihowtoimprove} &
%   S &
%   2.7M &
%   \xmark &
%   WikiHow &
%   \xmark &
%   \xmark &
%   % \xmark &
%   / \\ %\hline
% % Towards Modeling Revision Requirements in   wikiHow Instructions &
% \citet{bhat-etal-2020-towards} &
%   S &
%   6.95M &
%   \xmark &
%   WikiHow &
%   \xmark &
%   \xmark &
%   % \xmark &
%   \makecell[l]{Unrevised sentences \\ included} \\ \hline
% Understanding Iterative Revision from   Human-Written Text &
\citet{du-etal-2022-understanding-iterative} &
  S, P &
  31K* &
  \cmark &
  \begin{tabular}[c]{@{}l@{}}ArXiv\\      Wikipedia\\      WikiNews\end{tabular} &
  \cmark &
  \cmark &
  % \xmark &
  \makecell[l]{Across domain; \\ Revision history} \\ % \hline
% Read, Revise, Repeat: A System Demonstration for   Human-in-the-loop Iterative Text Revision &
\citet{du-etal-2022-read} &
  S, P &
  367K  &
  \cmark &
  \begin{tabular}[c]{@{}l@{}}ArXiv\\      Wikipedia\\      WikiNews\end{tabular} &
  \cmark &
  \cmark &
  % \xmark &
  \makecell[l]{Across domain; \\ Revision history} \\ % \hline
% Improving Iterative Text Revision by Learning   Where to Edit from Other Revision Tasks &
\citet{kim-etal-2022-improving} &
  S, P &
    367K &
  \cmark &
  \begin{tabular}[c]{@{}l@{}}ArXiv\\      Wikipedia\\      WikiNews\end{tabular} &
  \cmark &
  \cmark &
  % \xmark &
  \makecell[l]{Across domain; \\ Revision history; \\Including edit   tagging} \\ \hline

  % Modeling the Relationship between User Comments   and Edits in Doc. Revision &
\citet{zhang-etal-2019-modeling} &
  Sec &
  786K &
  \xmark &
  Wikipedia &
  \xmark &
  \cmark &
  % \xmark &
  \makecell[l]{Revision history; \\ Comment} \\ \hline
  % One Doc., Many Revisions: A Dataset for   Classification and Description of Edit Intents &
\citet{rajagopal-etal-2022-one} &
  D &
  9.3K &
  \cmark &
  Wikipedia &
  \cmark &
  \xmark &
  % \xmark &
  \makecell[l]{Revision History; \\ Comment} \\ \hline
  % SWIPE: A Dataset for Doc.-Level   Simplification of Wikipedia Pages &
\citet{laban-etal-2023-swipe} &
  D &
  145K* &
  \cmark &
  Wikipedia &
  \cmark &
  \cmark &
  % \xmark &
  / \\ \hline
    % Verba Volant, Scripta Volant: Understanding Post-publication Title Changes in News Outlets & 
  % \citet{Guo2022} &
  % S &	
  % 41.9K &	
  % \xmark &	
  % US News agencis &
  % \xmark &	
  % \xmark & 
  % Revision history \\ \hline
% NewsEdits: A News Article Revision Dataset and a   Doc.-Level Reasoning Challenge &
\citet{spangher-etal-2022-newsedits} &
  D &
  40M &
  \xmark &
  News &
  \xmark &
  \cmark &
  % \xmark &
  Multilingual \\ \hline
% ARXIVEDITS: Understanding the Human Revision   Process in Scientific Writing &
\citet{jiang-etal-2022-arxivedits} &
  S, D &
  13K &
  \cmark &
  ArXiv &
  \cmark &
  \cmark &
  % \cmark &
  / \\ \hline
  % Towards Automated Document Revision: Grammatical Error Correction, Fluency Edits, and Beyond &
\citet{mita-etal-2024-towards} &
  D, P &	
  64* &	
  \cmark & 
  ACL anthology &	
  \cmark &	
  \xmark	& 
  % \xmark &
  Comment \\ \hline

% Re3: A Holistic Framework and Dataset for   Modeling Collaborative Doc. Revision &
\citet{ruan-etal-2024-re3} &
  \makecell[l]{Sec, P, \\S, SS} &
  11.6K &
  \cmark &
  F1000Research &
  \cmark &
  \cmark &
  % \xmark &
  \makecell[l]{Review\&response \\  included} 
  
  \\ \Xhline{1pt}
\end{tabular}
\caption{Text-revision datasets. \textbf{Gran.} denotes revision granularity: S (Sentence), SS (Subsentence), P (Paragraph), D (Document), and Sec (Section).
\textbf{\#Pairs} reports the number of revised sentence pairs.
\textbf{Human} indicates the presence of human-annotated labels.
\textbf{Source} specifies the origin of the corpus.
\textbf{Intent} and \textbf{Action} indicate whether edit intentions and edit actions are annotated.
\textbf{Features} lists additional information included in each dataset. Note: (1) $\ast$ indicates the number of revised document pairs; (2) datasets grouped without separating lines correspond to reused and expanded corpora in follow-up work.}
%\caption{The publicly released datasets of text revision. \textbf{Gran.} indicates revision granularity, including S (Sentence), SS (Subsentence), P (Paragraph), D (Document), and Sec (Section). \textbf{\# pairs} reports the number of revised sentence pairs in the dataset. \textbf{Human} indicates whether the dataset contains human-annotated labels. \textbf{Source} indicates where the corpus comes from. \textbf{Intent} and \textbf{Action} indicate whether the dataset is annotated with edit intention and edit action. \textbf{Features} shows the other information or features included in the dataset. Note: (1) * indicates the number of revised document pairs; (2) the corpora of the papers in a group (no delimiting line between them) are reused and expanded in the follow-up work. }
\label{tab:datasets-all}
\end{table*}

\section{An Example of Workflow}
\label{sec:example_workflow}
\begin{figure*}[t]
    \centering
    \includegraphics[width=1\linewidth]{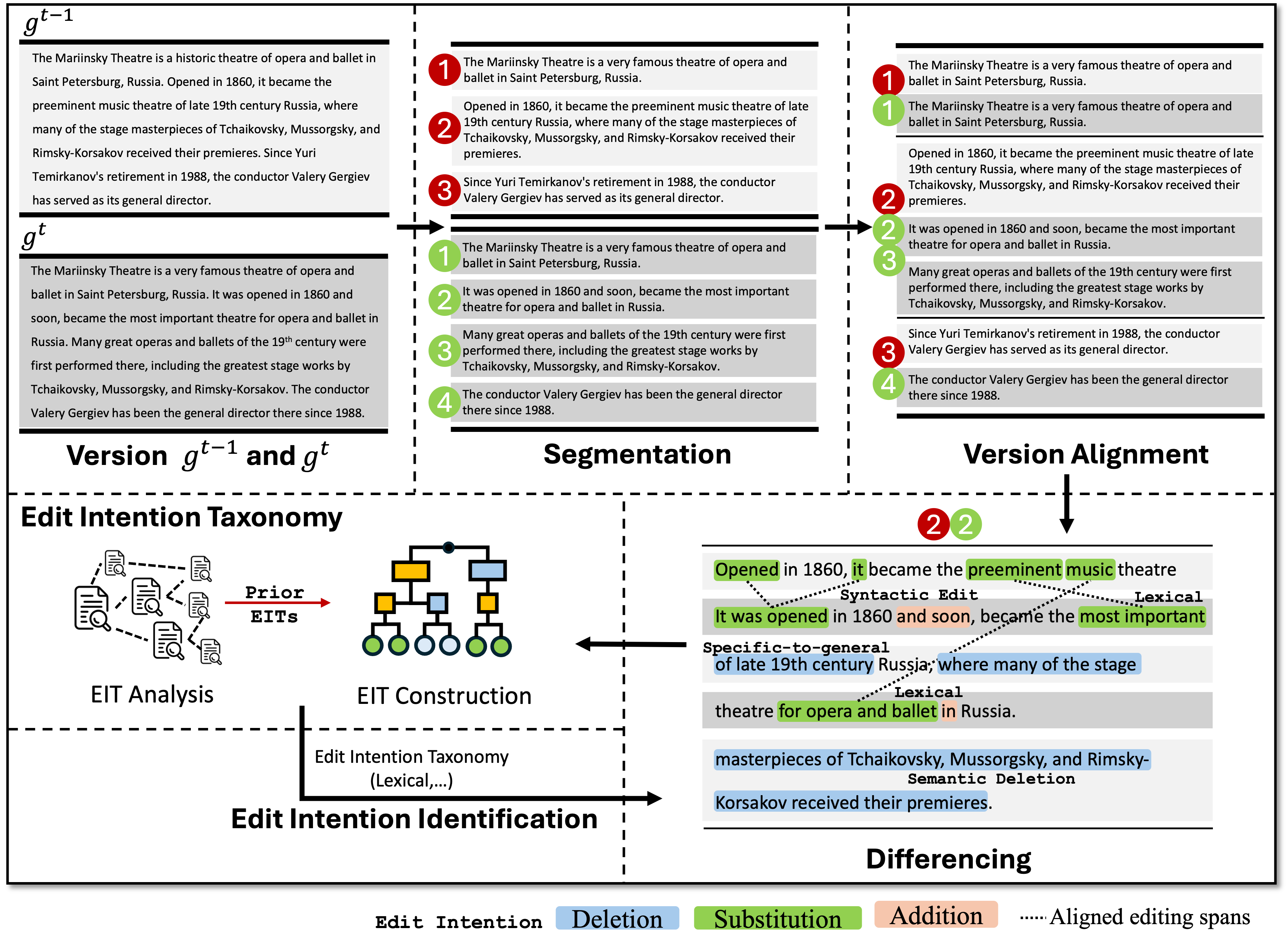}
    \caption{An example of the workflow of revision dataset construction.}
    \label{fig:edit_workflow_complete}
\end{figure*}

We give a concrete example of revision dataset construction in a desired scenario, i.e., every step works well, in Figure \ref{fig:edit_workflow_complete}. Given two paragraphs of the same Wikipedia article - Mariinsky Theater, segment them into sentences. The $P^{t-1}$ and $P^{t}$ are divided into 3 and 4 sentences, respectively. Then, we do a Cartesian product of sentences from the two versions, and apply the version alignment algorithm to find correspondence, resulting in 3 matches, i.e., 1 to 1, 2 to 2\&3, and 3 to 4. The match of 2 to 2\&3 indicates a finer edit action of split, while the other two matches are revised. Using 2 to 2 as a demonstration for differencing, we highlight the results of the differencing algorithm with green (Substitution), orange (Addition), and blue (Deletion) blocks. The dashed lines connect the editing spans that indicate one substantial edit. Then, we label those edits with an EIT.

\section{EIT Characteristics}

\subsection{Complete EITs Statistics}
\label{sec:eit_stats}
% Table for EIT statistics
% \usepackage{multirow}
\begin{table*}[ht]
\centering
% \vspace{-1em}
\small
\begin{tabular}{l|l|p{1.2cm}|p{0.3cm}|lll|p{0.6cm}p{0.8cm}|p{0.8cm}}
% \hhline{|==|==|==|==|==|==|==|==|==|}
\Xhline{1.5pt}
\multirow{2}{*}{\textbf{alias}} &
  \multirow{2}{*}{\textbf{paper}} &
  \multirow{2}{*}{\textbf{domain}} &
  \multirow{2}{*}{\textbf{\#le.}} &
  \multicolumn{3}{c|}{\textbf{has\_definition}} &
  \multicolumn{2}{c|}{\textbf{has\_examples}} &
  \multirow{2}{*}{\textbf{link?}} \\ \cline{5-9}
 &
   &
   &
   &
  \multicolumn{1}{l|}{\textbf{l1}} &
  \multicolumn{1}{l|}{\textbf{l2}} &
  \textbf{l3} &
  \multicolumn{1}{l|}{\textbf{extents}} &
  \textbf{from} &
   \\ \Xhline{1.5pt}
Fai81 &
  \citealp{Faigley1981} &
  Writings &
  3 &
  \multicolumn{1}{l|}{Yes} &
  \multicolumn{1}{l|}{Yes} &
  Partial &
  \multicolumn{1}{l|}{Partial} &
  Inside &
  / \\ \hline

Zha15 &
  \citealp{zhang-litman-2015-annotation} &
  Writings &
  3 &
  \multicolumn{1}{l|}{Yes} &
  \multicolumn{1}{l|}{Yes} &
  No &
  \multicolumn{1}{l|}{Full} &
  Inside &
  /  \\ \hline
Zha16(A) &
    \citealp{zhang-etal-2016-argrewrite} &
  Writings &
  2 &
  \multicolumn{1}{l|}{Yes} &
  \multicolumn{1}{l|}{Yes} &
   &
  \multicolumn{1}{l|}{Full} &
  Outside &
  Active \\ \hline
Zha17 &
  \citealp{zhang-etal-2017-corpus} &
  Writings &
  2 &
  \multicolumn{1}{l|}{Yes} &
  \multicolumn{1}{l|}{Yes} &
  / &
  \multicolumn{1}{l|}{Full} &
  Both &
  Active \\ \hline
Fal21 &
    \citealp{faltings-etal-2021-text} &
  Writings, Wikipedia &
  2 &
  \multicolumn{1}{l|}{Yes} &
  \multicolumn{1}{l|}{Yes} &
  / &
  \multicolumn{1}{l|}{Partial} &
  Inside &
  Broken \\ \hline
Du22 &
    \citealp{du-etal-2022-understanding-iterative} &
  Writings,   Wikipedia, News &
  2 &
  \multicolumn{1}{l|}{Yes} &
  \multicolumn{1}{l|}{Yes} &
  / &
  \multicolumn{1}{l|}{Full} &
  Both &
  Active \\ \hline
Jia22 &
    \citealp{jiang-etal-2022-arxivedits} &
  Writings &
  2 &
  \multicolumn{1}{l|}{Yes} &
  \multicolumn{1}{l|}{Yes} &
  / &
  \multicolumn{1}{l|}{Full} &
  Both &
  Active \\ \hline
Kas22 &
    \citealp{Kashefi2022-xs} & 
  Writings &
  2
   &
  \multicolumn{1}{l|}{Yes} &
  \multicolumn{1}{l|}{Yes} &
  /
   &
  \multicolumn{1}{l|}{Full} &
  Both
   &
   Active
   \\ \hline
   Rua24&
 \citealp{ruan-etal-2024-re3}&
  Writings &
  2 &
  \multicolumn{1}{l|}{Yes} &
  \multicolumn{1}{l|}{Yes} &
  / &
  \multicolumn{1}{l|}{Full} &
  Both &
  Active \\ \hline

  Rat85 &
  \citealp{Rathjens1985} &
  Writings &
  1 &
  \multicolumn{1}{l|}{Yes} &
  \multicolumn{1}{l|}{/} &
  / &
  \multicolumn{1}{l|}{Full} &
  Inside &
  / \\ \hline
Zha16(B) &
    \citealp{zhang-litman-2016-using} &
  Writings &
  1 &
  \multicolumn{1}{l|}{Yes} &
  \multicolumn{1}{l|}{/} &
  / &
  \multicolumn{1}{l|}{Partial} &
  Inside &
  /
   \\ \Xhline{1pt}
Yan16 &
\citealp{Yang_Halfaker_Kraut_Hovy_2016} &
  Wikipedia &
  3 &
  \multicolumn{1}{l|}{Yes} &
  \multicolumn{1}{l|}{Yes} &
  No &
  \multicolumn{1}{l|}{No} &
  / &
  /  \\ \hline
Jon08 &
\citealp{Jones2008} &
  Wikipedia &
  2 &
  \multicolumn{1}{l|}{Yes} &
  \multicolumn{1}{l|}{No} &
  / &
  \multicolumn{1}{l|}{No} &
  / &
  /  \\ \hline
Dax12 &
\citealp{daxenberger-gurevych-2012-corpus} &
  Wikipedia &
  2 &
  \multicolumn{1}{l|}{Yes} &
  \multicolumn{1}{l|}{Yes} &
  / &
  \multicolumn{1}{l|}{Full} &
  Inside &
  Moved \\ \hline
Dax13 &
\citealp{daxenberger-gurevych-2013-automatically} &
  Wikipedia &
  2 &
  \multicolumn{1}{l|}{Yes} &
  \multicolumn{1}{l|}{Yes} &
  / &
  \multicolumn{1}{l|}{No} &
  / &
  Broken \\ \hline

Yan17 &
\citealp{yang-etal-2017-identifying-semantic} &
  Wikipedia &
  2 &
  \multicolumn{1}{l|}{Yes} &
  \multicolumn{1}{l|}{Yes} &
  / &
  \multicolumn{1}{l|}{Full} &
  Outside &
  Active  \\ \hline 
Lab23 &
\citealp{laban-etal-2023-swipe} &
  Wikipedia &
  2 &
  \multicolumn{1}{l|}{Yes} &
  \multicolumn{1}{l|}{Yes} &
  / &
  \multicolumn{1}{l|}{Full} &
  Both &
  Active \\ \hline
Pfe06 &
\citealp{Pfeil2006} &
  Wikipedia &
  1 &
  \multicolumn{1}{l|}{Yes} &
  \multicolumn{1}{l|}{/} &
  / &
  \multicolumn{1}{l|}{Full} &
  Inside &
  / \\ \hline
Liu11 &
\citealp{Liu2011} &
  Wikipedia &
  1 &
  \multicolumn{1}{l|}{Yes} &
  \multicolumn{1}{l|}{/} &
  / &
  \multicolumn{1}{l|}{No} &
  / &
  / \\ \hline
Far18 &
\citealp{faruqui-etal-2018-wikiatomicedits} &
  Wikipedia &
  1 &
  \multicolumn{1}{l|}{Yes} &
  \multicolumn{1}{l|}{/} &
  / &
  \multicolumn{1}{l|}{Full} &
  Both & 
  Active
   \\ \hline
Raj22 &
\citealp{rajagopal-etal-2022-one} &
  Wikipedia &
  1 &
  \multicolumn{1}{l|}{Yes} &
  \multicolumn{1}{l|}{/} &
  / &
  \multicolumn{1}{l|}{Full} &
  Both &
  Active \\ \Xhline{1pt}
Ant20 &
\citealp{anthonio-etal-2020-wikihowtoimprove} &
  WikiHow &
  1 &
  \multicolumn{1}{l|}{No} &
  \multicolumn{1}{l|}{/} &
  / &
  \multicolumn{1}{l|}{Full} &
  Both &
  Active \\  \Xhline{1pt}
Spa22 &
\citealp{spangher-etal-2022-newsedits} &
  News &
  1 &
  \multicolumn{1}{l|}{Yes} &
  \multicolumn{1}{l|}{/} &
  / &
  \multicolumn{1}{l|}{Full} &
  Both &
  Active \\ \hline
Guo22 &
\citealp{Guo2022} &
  News &
  1 &
  \multicolumn{1}{l|}{Yes} &
  \multicolumn{1}{l|}{/} &
  / &
  \multicolumn{1}{l|}{Full} &
  Inside &
  Active \\ \Xhline{1.5pt}
\end{tabular}
% \vspace{-0.8em}
\caption{It shows the application \textbf{domain} of the taxonomy in a \textbf{paper}; \textbf{\#le.} is the number of levels and whether it is flat (\textbf{\#le.}=1) or hierarchical (\textbf{\#le.}>1); whether the categories in level \# (\textbf{l\#}) has definitions (\textbf{has\_definition}); whether `Full', `Partial' or `No' leaf categories (\textbf{extents}) have concrete revision examples (\textbf{has\_examples}) and they are \textbf{from} `Inside' of the paper, `Outside' of the paper through provided link (\textbf{link?}) or `Both'. `/' denotes not applicable.}
\label{tab:paper-list}
% \vspace{-1.5em}
\end{table*}
Table \ref{tab:paper-list} summarizes the characteristics. 

\subsection{Application Domains and Availability}
\label{sec:edit_domains}

% \noindent\textbf{Application Domains}
% EITs have been developed across diverse application domains.
% A substantial portion of the literature focuses on Wikipedia, examining collaborative behaviors, editor roles, and semantic edit intentions
% \cite{Pfeil2006, Jones2008, daxenberger-gurevych-2012-corpus, Yang_Halfaker_Kraut_Hovy_2016, yang-etal-2017-identifying-semantic}.
% Writing-focused studies analyze revisions in student, technical, and academic texts to understand revision intent, clarity, and argumentative development
% \cite{Faigley1981, Rathjens1985, zhang-litman-2015-annotation, Kashefi2022-xs}.
% Other work targets news articles and headlines
% \cite{spangher-etal-2022-newsedits, Guo2022}
% or instructional texts such as wikiHow
% \cite{anthonio-etal-2020-wikihowtoimprove}.
% More recent EITs aim to generalize across multiple domains
% \cite{du-etal-2022-understanding-iterative}.

\noindent\textbf{Data Sources and Resource Availability}
Prior work varies considerably in the availability of publicly released datasets, annotation guidelines, and source code.
Some studies release substantial resources, including annotated corpora and preprocessing pipelines
\cite{daxenberger-gurevych-2012-corpus, yang-etal-2017-identifying-semantic, du-etal-2022-understanding-iterative},
while others provide datasets without aligned taxonomies
\cite{Guo2022}
or release trained models and crawlers
\cite{anthonio-etal-2020-wikihowtoimprove}.
However, broken or outdated links remain common
\citealp{daxenberger-gurevych-2013-automatically},
limiting reproducibility and reuse.

% Overall, existing EITs exhibit substantial conceptual overlap but differ in structure, definition quality, domain coverage, and resource support.

\subsection{Observations on EIT Lineage}
\label{sec:lineage}

Fai81\footnote{For compactness, we use aliases for the papers to reference their EITs. See Table~\ref{tab:paper-list} for full reference.} is the earliest EIT in our survey. Jon08 extends it by simplifying Macro/Micro-structure subcategories and adding Wikipedia-specific categories. Dax12 and Zha15 both inherit the high-level split of Surface Changes vs.\ Text-Base Changes from Fai81, while adapting subcategories to their respective settings: Dax12 emphasizes Wikipedia editing objects (e.g., file, template), and Zha15 targets structural elements of academic writing (e.g., claims/ideas, introductory materials).

Du22 reframes the taxonomy around meaning impact via Meaning-changed vs.\ Non-meaning-changed, defining Non-meaning-changed subcategories by building on Raj85 and \citet{Harris2017-wa}. In parallel, Liu11 compresses Pfe06’s detailed scheme to improve feasibility for automatic classification, and further augments Pfe06 by introducing references-related categories alongside its link-related ones, motivated by their role in article quality.

Zha16(B) simplifies Zha15 by collapsing Surface subcategories into a single Surface category to focus on augmentative changes, and merges Rebuttal into Warrant based on the rarity of rebuttal edits observed in Zha15. Yan16 extends Dax12 by further fine-graining Reference, using edit intentions to identify editor roles, whereas Dax12 uses them to study collaborative behavior.

Yan17 distills and extends categories from Fai81, Dax12, and Zha16(B), organizing top-level categories by whether revisions are general or Wikipedia-specific, and introducing many new subcategories. Fal21 then reorganizes Yan17’s leaf categories into Fluency, Content, and Other, distinguishing grammar/structure edits from meaning-changing edits. Raj22 further refines Yan17 by collapsing similar categories and merging Wikipedia-specific categories into Other.

Finally, Guo22 draws on Yan17 and Dax13 but tailors the taxonomy to news headline revisions, replacing unsuitable categories and adding journalism-driven ones \cite{ClickBait, SixThingYouDidntKnowBboutHeadingWriting}. Ant20 is derived from manual inspection of a sampled set of sentence revision pairs.

\section{Annotation Evaluation Metrics}
\label{sec:eval_metrics}
We summarize the evaluation metrics commonly used to assess the performance of edit intention identification models under these different formulations and annotation settings.

In multi-label classification, each instance (here, an edit) can belong to multiple labels. We consider example-based and label-based evaluation. The key difference between them is what gets averaged. \textit{Example-based Metrics} are weighting each edit equally for multi-label classification. It can evaluate the overall usefulness on real instances. We denote the set of relevant categories for each edit $e_i \in E$ as $y_i \in C$ and the set of predicted categories as $h(e_i)$. The accuracy of a multi-label classifier is defined as
\begin{equation}
    ACC = \frac{1}{|E|}\sum^{|E|}_{i}\frac{|h(e_{i})\cap y_{i}|}{h(e_{i}\cup y_{i})} 
\end{equation}
which corresponds to the Jaccard similarity of $h(e_i)$ and $y_i$ averaged over all edits. Example-based precision and recall are defined as 
% \begin{equation}
%     \begin{aligned}
%         P & = \frac{1}{|E|}\sum^{|E|}_{i}\frac{|h(e_{i})\cap y_{i}|}{h(e_{i})}  \\
%         R & = \frac{1}{|E|}\sum^{|E|}_{i}\frac{|h(e_{i})\cap y_{i}|}{y_{i}} \\
%         F1 & = \frac{1}{|E|}\sum^{|E|}_{i}\frac{2 \times |h(e_{i})\cap y_{i}|}{|h(e_{i})|+|y_{i}|}
%     \end{aligned}
% \end{equation}
\begin{equation}
    P  = \frac{1}{|E|}\sum^{|E|}_{i}\frac{|h(e_{i})\cap y_{i}|}{h(e_{i})}  
\end{equation}

\begin{equation}
    R = \frac{1}{|E|}\sum^{|E|}_{i}\frac{|h(e_{i})\cap y_{i}|}{y_{i}}
\end{equation}

\begin{equation}
    F1 = \frac{1}{|E|}\sum^{|E|}_{i}\frac{2 \times |h(e_{i})\cap y_{i}|}{|h(e_{i})|+|y_{i}|}
\end{equation}

We report macro- and micro-averaged F1 scores for label-based measures for \textit{Label-based metrics}, which report robustness across categories and imbalance. It treats each category equally (macro) or each label occurrence equally (micro). 

\textit{Exact Match} evaluates whether the predicted labels are the same as the actual labels, calculated as 

\begin{equation}
    ACC_{exact} = \frac{1}{|E|} \sum^{|E|}_{i=1}I
\end{equation}
where $I=1$ if $h(e_{i}) = y_{i}$ and $I=0$ otherwise.

% Two tables for data sources and links section
\begin{table*}[ht]
\centering
\small
\begin{tabular}{l|p{0.8cm}|p{0.8cm}|p{7.7cm}|p{1cm}}
\hline
\textbf{paper} & \textbf{code?} & \textbf{data?} & \textbf{sources} & \textbf{link?} \\ \hline
Fai81    & No   & No & (In)experienced   student and expert revisions                & No    \\ \hline
Zha15    & No   & No & Student written   papers                                      & No    \\ \hline
Zha16(A) & No   & No         & Student writings                                              & Yes   \\ \hline
Zha17    & No   & Ye   & Student writings                                              & Yes   \\ \hline
Fal21    & Yes  & Ye   & Wikipedia revision   histories                                & Yes   \\ \hline
Du22     & Yes  & Ye   & Formally   human-written text(Wikipedia, ArXiv, Wikinews)     & Yes   \\ \hline
Jia22    & Yes  & Ye   & ArXiv Papers   Revisions                                      & Yes   \\ \hline
Kas22    & /    & Ye   & argumentative writing   essays                                & Yes   \\ \hline
Rua24 &
  Yes &
  Ye &
  F1000RD   dataset in Kuznetsov et al., 2022, ARR-22 subset of the NLPeer corpus in   Dycke et al., 2023 &
  Yes \\ \hline
Rat85    & /    & /          & /                                                             & No    \\ \hline
Zha16(B) & No   & No & High school student   written papers                          & No    \\ \hline
Yan16    & No   & No & Three datasets from   English edition of Wikipedian           & No    \\ \hline
Jon08    & No   & No & (Non-)featured   Wikipedia articles revision histories        & No    \\ \hline
Dax12    & No   & Ye   & (Non-)featured   Wikipedia articles revision histories        & Yes   \\ \hline
Dax13    & No   & Ye   & Wikipedia revision   histories                                & Yes   \\ \hline
Yan17    & Yes  & Ye   & Wikipedia revision   histories                                & Yes   \\ \hline
Lab23    & Yes  & Ye   & Wikipedia’s revision   history                                & Yes   \\ \hline
Pfe06    & No   & No & Wikipedia page   revision histories                           & No    \\ \hline
Liu11    & No   & No & Different   quality-level Wikipedia articles based on quality & No    \\ \hline
Far18    & No   & Yes   & revision histories                                            & Yes   \\ \hline
Raj22    & No   & Yes   & Wikipedia revision   histories                                & Yes   \\ \hline
Ant20    & Yes  & Yes   & wikiHow   (non-)featured articles revision histories          & Yes   \\ \hline
Spa22    & Yes  & Yes   & News revision   histories                                     & Yes   \\ \hline
Guo22    & No   & Yes   & News headlines from   major US news agencies                  & Yes   \\ \hline
\end{tabular}
\vspace{-0.8em}
\caption{This table summarizes whether the literature provided their code and data. \textbf{code?} indicates if the code for preprocessing or regenerating data was released. \textbf{data?} shows if the data was provided along with their taxonomy. \textbf{sources} specifies the sources of the revisions their taxonomy and analysis are based on. \textbf{link?} indicates if the links to their code or data were provided. `/' denotes not applicable.}
\vspace{-1em}
\label{tab:paper-resources}
\end{table*}

% \caption{This table summarizes whether the literature provided their code and data. \textbf{code?} indicates if the code for preprocessing or regenerating data was released. \textbf{data?} shows if the data was provided along with their taxonomy. \textbf{sources} specifies the sources of the revisions their taxonomy and analysis are based on. \textbf{link?} indicates if the links to their code or data were provided. `Moved' means the source link has changed but is still accessible. `Broken' indicates the provided link is not accessible and the source cannot be found via searches. `/' denotes not applicable.}
% \input{sections/tables/datacode-links}

\section{Demonstration: Differencing Algorithms}
\label{sec:demo_diff_alg}

\begin{figure}
    \centering
    \includegraphics[width=1\linewidth]{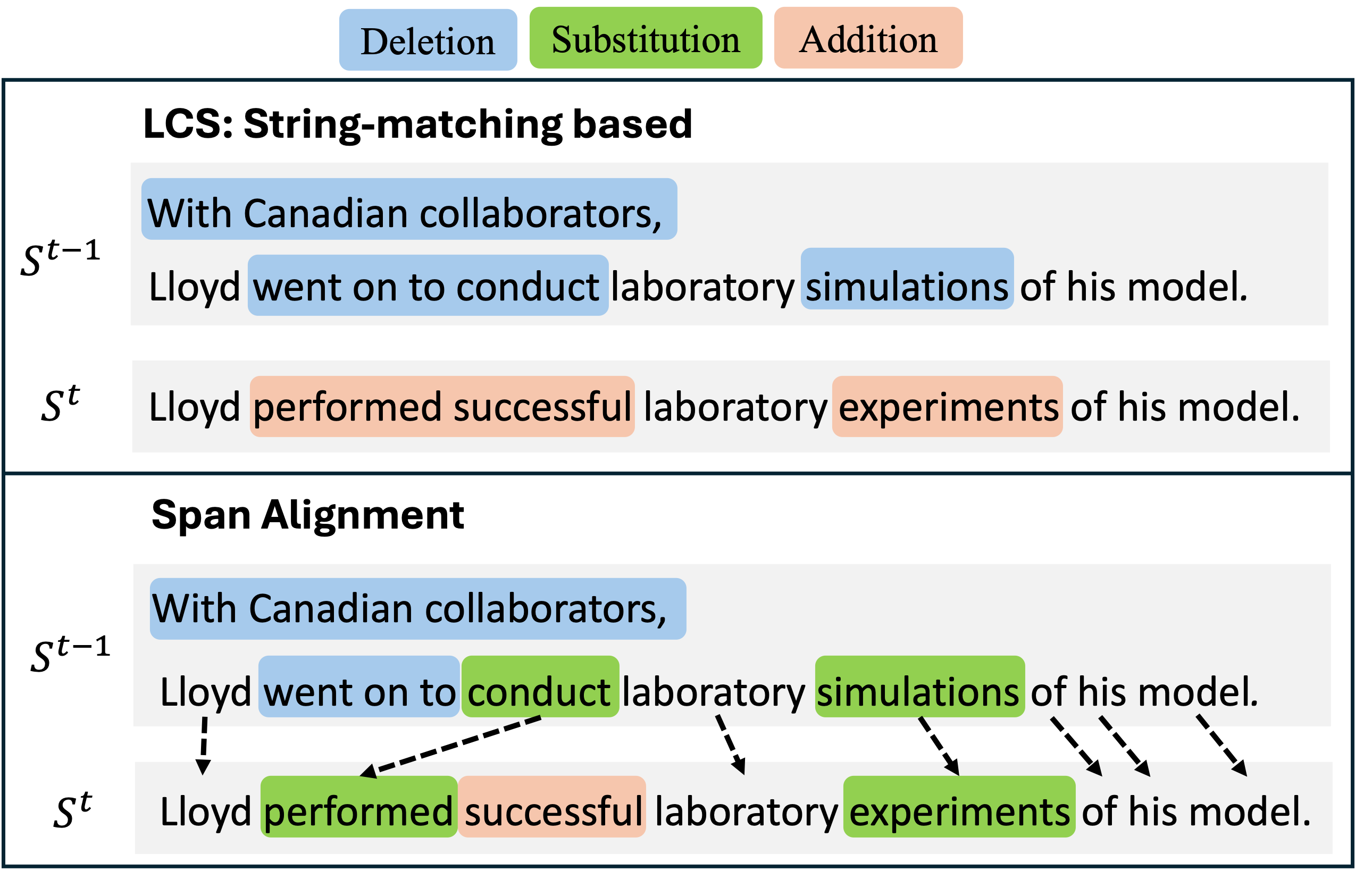}
    \caption{Comparison of string-matching–based differencing and semantic span alignment. While LCS-based differencing highlights surface-level changes, span alignment captures semantically meaningful edit operations such as deletions, substitutions, and additions.}
    \label{fig:differencing_demo}
\end{figure}

Figure \ref{fig:differencing_demo} demonstrates the results of the string-matching-based LCS algorithm and semantic-aware span alignment. For example, deleting the phrase ``went on to conduct'' should ideally be decomposed into deleting ``went on to'' and substituting ``conduct'' with ``performed'', yet LCS-based differencing treats this as a single deletion. Similarly, semantically equivalent substitutions such as ``conduct'' $\leftrightarrow$ ``performed'' or ``simulations'' $\leftrightarrow$ ``experiments'' are not explicitly aligned. This behavior mixes deletions, substitutions, and additions at the string level, making subsequent edit alignment and edit intention identification difficult. Unlike LCS-based differencing, the span align algorithm captures semantic substitutions (e.g., “conduct” $\leftrightarrow$ “performed” in Figure 3) and directly aligns corresponding edit spans.

\section{The Use of Large Language Model}
To enhance readability, we employed OpenAI GPT-5.2 strictly as a language editing tool for grammar
correction and stylistic refinement. 
Its use was limited to functions analogous to conventional
proofreading and did not contribute to the conception, methodology, analysis, or scientific content of
this work.
%%%%%%%%%%%%%%%% ACL2026 version end %%%%%%%%%%%%%%%

\end{document}